\documentclass[format=acmsmall, review=false, screen=true]{acmart}

\usepackage[T1]{fontenc} 
\usepackage[utf8]{inputenc}
\usepackage{amsmath}
\usepackage{bm} 
\usepackage{graphicx}
\usepackage{tikz}
\usepackage{tcolorbox}
\usepackage{pdflscape}
\usepackage{rotating}
\usepackage{xcolor}
\usepackage{fancyhdr} 
\usepackage{color, colortbl}
\usepackage{longtable}
\usepackage{array}
\usepackage{multirow}
\usepackage{graphicx}
\usepackage{enumitem}
\usepackage{fontawesome}
\usepackage{multirow}
\usepackage{balance}
\usepackage{tablefootnote}
\usepackage{colortbl}
\tcbuselibrary{skins}
\usepackage{hyperref}
\usepackage[export]{adjustbox}
\usepackage{framed}
\usepackage{arydshln}

\usepackage{subfig}
\usepackage{xcolor}

\usepackage{makecell}
\usepackage{threeparttable}

\usepackage{indentfirst}
\usepackage{listings}
\lstdefinelanguage{Kotlin}{
  morekeywords={val, var, fun, if, else, while, for, in, return, open, class, @Nullable},
  sensitive=true,
  morecomment=[l]{//},
  morecomment=[s]{/*}{*/},
  morestring=[b]",
  morestring=[s]{"""*}{*"""},
}

\definecolor{darkred}{rgb}{0.6680, 0.3203, 0.3789}
\definecolor{darkgreen}{rgb}{0.0664, 0.3867, 0.1602}
\definecolor{lightred}{rgb}{1.0, 0.9180, 0.9101}
\definecolor{lightgreen}{rgb}{0.8516, 0.9180, 0.8789}
\lstdefinelanguage{diff}{
  sensitive=true,
}
\lstdefinestyle{gitpatch}{
    backgroundcolor=\color{white},
    basicstyle=\ttfamily,
    keywordstyle=\color{green},
    stringstyle=\color{red},
    commentstyle=\color{black},
    morecomment=[l][\color{darkred}]{-},  
    morecomment=[l][\color{darkgreen}]{+}, 
    morecomment=[s]{@@}{@@},
    showstringspaces=false,
    tabsize=2,
}

\newcommand{\toolName}{\textsc{DEVLoRe}}
\newcommand{\dforj}{Defects4J}
\newcommand{\leftquote}{``}
\newcommand{\rightquote}{''}

\setcopyright{acmlicensed}
\acmJournal{TOSEM}
\copyrightyear{2025}
\acmYear{2025}
\acmDOI{XXXXXXX.XXXXXXX}

\acmISBN{978-1-4503-XXXX-X/18/06}

\begin{document}

\title{Integrating Various Software Artifacts for Better LLM-based Bug Localization and Program Repair}

\author{Qiong Feng}
\affiliation{%
  \institution{Nanjing University of Science and Technology; State Key Lab. for Novel Software Technology, Nanjing University}
  \city{Nanjing}
  \country{China}
}
\email{qiongfeng@njust.edu.cn}

\author{Xiaotian Ma}
\affiliation{%
  \institution{Nanjing University of Science and Technology}
  \city{Nanjing}
  \country{China}
}
\email{xyzboom@njust.edu.cn}

\author{Jiayi Sheng}
\affiliation{%
  \institution{Nanjing University of Science and Technology}
  \city{Nanjing}
  \country{China}
}
\email{shengjiayi@njust.edu.cn}

\author{Ziyuan Feng}
\affiliation{%
  \institution{Nanjing University of Science and Technology}
  \city{Nanjing}
  \country{China}
}
\email{azumaseren@njust.edu.cn}

\author{Wei Song}
\affiliation{%
  \institution{Nanjing University of Science and Technology}
  \city{Nanjing}
  \country{China}
}
\email{wsong@njust.edu.cn}

\author{Peng Liang}
\authornote{Corresponding author}
\affiliation{%
  \institution{School of Computer Science, Wuhan University}
  \city{Wuhan}
  \country{China}
}
\email{liangp@whu.edu.cn}


\begin{abstract}

LLMs have garnered considerable attention for their potential to streamline Automated Program Repair (APR). LLM-based approaches can either insert the correct code using an infilling-style technique or directly generate patches when provided with buggy methods, aiming for plausible patches to pass all tests. However, most of LLM-based APR methods rely on a single type of software information, such as issue descriptions or error stack traces, without fully leveraging a combination of diverse software artifacts. Human developers, in contrast, often use a range of information — such as debugging data, issue discussions, and error stack traces — to diagnose and fix bugs. Despite this, many LLM-based approaches do not explore which specific types of software information best assist in localizing and repairing software bugs. Addressing this gap is crucial for advancing LLM-based APR techniques.

To investigate this and mimic the way human developers fix bugs, we propose \toolName{} (short for \underline{DEV}eloper \underline{Lo}calization and \underline{Re}pair). In this framework, LLMs first use issue content (description and discussion) and stack error traces to localize buggy methods, then rely on debug information in buggy methods and issue content and stack error to localize buggy lines and generate valid patches. We evaluated the effectiveness of issue content, error stack traces, and debugging information in bug localization and automatic program repair. Our results show that while issue content and error stack is particularly effective in assisting LLMs with fault localization and program repair respectively, different types of software artifacts complement each other in addressing various bugs. By incorporating these three types of artifacts and using the Defects4J v2.0 dataset for evaluation, \toolName{} successfully localizes 49.3\% of single-method bugs and generates 56.0\% plausible patches. Additionally, \toolName{} can localize 47.6\% of non-single-method bugs and generates 14.5\% plausible patches. Moreover, our framework streamlines the end-to-end process from buggy source code to a complete repair, and achieves a 39.7\% and 17.1\% of single-method and non-single-method bug repair rate, outperforming current state-of-the-art APR methods. \textcolor{black}{Furthermore, we re-implemented and evaluated our framework, demonstrating its effectiveness in resolving 9 unique issues compared to other state-of-the-art frameworks using the same or more advanced models on SWE-bench Lite. We also discussed whether a leading framework for Python code can be directly applied to Java code, or vice versa.} The source code and experimental results of this work for replication are available at \url{https://github.com/XYZboom/DEVLoRe}.

\end{abstract}




\maketitle

\section{Introduction}
Automatic Program Repair (APR) streamlines the process of identifying and correcting code defects, significantly reducing the time and effort required for manual bug fixing~\cite{le:2019apr}. Traditional APR techniques employ various methods, including template-based approaches~\cite{template:ghanbari_practical_2019,template:le_goues_genprog_2011,template:le_history_2016,template:liu_tbar_2019,template:long_staged_2015,template:mechtaev_angelix_2016} and neural machine translation (NMT)~\cite{nmt:chen_sequencer_2019,nmt:jiang_cure_2021,nmt:li_dlfix_2020,nmt:zhu_syntax-guided_2021}, to generate potential patches that are both syntactically valid and semantically meaningful. Although these methods can produce correct patches for certain bugs, they also have notable limitations. NMT models rely heavily on bug-fixing training data, making them unable to generate patches for new or unseen types of bugs. On the other hand, template-based approaches suffer from a limited set of templates and struggle to address more complex, nontrivial bug fixes~\cite{xia_automated_2023}.

Recent studies have explored the use of LLMs for APR, either by having LLMs fill in the correct code in buggy methods using an infilling-style technique or by directly generating patches when provided with buggy methods~\cite{chatrepair,xia_agentless_2024,xia_fuzz4all_2024,li_giantrepair_2024,xia_automated_2023,fitrepair,jiang_impact_2023,repilot,zhang_gamma_2023,silva2023repairllama}. The initial results demonstrate the ability of LLMs to correctly repair real-world bugs, including those that were previously unrepairable by existing APR approaches. One of the best-performing frameworks is \texttt{Agentless}~\cite{xia_agentless_2024}, which feeds an LLM with only issue description and can automatically solve GitHub issues. These promising outcomes highlight the potential of LLMs to develop more effective APR methods. 

However, there are still two major \textbf{limitations} that need to be addressed:

\begin{itemize} 
\item Current LLM-based approaches do not fully incorporate various types of software artifacts. Most LLMs rely on just one or two kinds of software artifacts, such as issue descriptions or code structure, while other important artifacts are underutilized. For instance, debugging information, which is a critical tool for human developers in diagnosing and resolving bugs, is often not fully leveraged. 

\item While various LLM-based approaches make use of different software artifacts, such as issue descriptions and error stack traces, it remains unclear which specific type of information most effectively aids LLMs in localizing and automatically repairing software bugs. 
\end{itemize}

To address these two limitations and further explore the ability of LLMs to localize and fix software bugs (bug, defect, and fault are used interchangeably in this paper), we propose feeding LLMs different types of software artifacts to determine which information best leverages their capabilities in bug localizing and fixing. The rationale behind this approach is that human developers typically do not rely on a single type of information when localizing and fixing software bugs. Instead, they combine various sources of information in software development, such as issue descriptions, proof of concept (PoC), stack traces, discussions in issues, and more. Based on developers' experience, having more information helps to better understand the root cause of bugs, ultimately leading to more effective bug localization and fixes.

To achieve this, we propose the \toolName{} framework, which asks LLMs to mimic human developers for bug localization and program repair. Along with this framework, we design two tools to extract executed methods in failed test cases and debugging information of buggy methods. In this framework, we feed the chosen LLM with three types of software artifacts: issue content (including issue description and discussion), error stack trace, and debug information. Then, using the well-known Defects4J dataset~\cite{defects4j}, we evaluate how these different types of software artifacts contribute to effectively localizing and fixing bugs. Our experiment results demonstrate that issue content is the most effective indicator for buggy method's localization, achieving 43.6\% for localizing single-method bugs and 40.6\% for localizing non-single-method bugs. Stack error trace proves to be the best indicator for single-method bug fixing, with a precision of 27.0\%. Additionally, our findings highlight that different types of software information can complement each other in the process of localizing and fixing bugs. By combining issue content and error stack trace, we achieve a state-of-the-art performance of 49.3\% in localizing the single-method bugs. Furthermore, incorporating issue content, error stack trace, and debugging information results in a fix rate of 43.1\% for single-method bugs. 

As our \toolName{} framework does not specify that the localization and fix of buggy methods should be singular, it can localize and fix bugs across multiple methods (non-single-method bugs). For example, by combining all three software artifacts, \toolName{} can fix 9.4\% of non-single-method bugs with provided buggy locations. Moreover, similar to \texttt{Agentless}~\cite{xia_agentless_2024}, \toolName{} relies on LLMs throughout the entire end-to-end process of fault localization and program repair. It can successfully fix 28.0\% of single-method bugs and 11.2\% of non-single-method bugs when combining all three artifacts. To the best of our knowledge, our approach outperforms current state-of-the-art methods.

To summarize, our \textbf{contributions} in this paper are as follows:
\begin{enumerate}
    \item To the best of our knowledge, this work is the first to compare different software artifacts in assisting LLMs' ability to perform \textcolor{black}{both} fault localization and program repair. The results can help developers understand LLMs' potential when provided with a variety of software artifacts.
    
    \item We propose a simple and lightweight framework that leverages LLMs to conduct an end-to-end process for bug localization and program repair. Accompanying this framework we design a strict input/output prompt. The \toolName{} framework demonstrates a strong ability to localize and fix more software bugs in less time and at a lower cost, compared to current state-of-the-art methods \textcolor{black}{on the \dforj{} dataset}.

    \item We evaluated our framework on SWE-bench Lite and demonstrated \toolName{}'s effectiveness in resolving 9 unique issues compared to other state-of-the-art frameworks using the same or more advanced models. We further discussed whether a leading framework for Python code can be directly applied to Java code, or vice versa.
\end{enumerate}

The paper is structured as follows: Section~\ref{sec:motivation} discusses the motivation of using various software artifacts, Section~\ref{sec:approach} outlines the proposed approach, Section~\ref{sec:experiment} introduces the experiment setup and the research questions, Sections~\ref{sec:results} and~\ref{sec:discussion} present and discuss the results respectively, Section~\ref{sec:related} reviews the related work, and Section~\ref{sec:conclusion} concludes this work with future directions.
\section{Three Motivating Examples}
\label{sec:motivation}

Our assumption is that LLM-based program repair tools have the potential to be much more effective if they can leverage a variety of software artifacts, such as code snippets, version history, documentation, and even testing outputs. By providing LLMs with a rich set of contextual data, they can understand and localize bugs more accurately and offer more precise fixes. This assumption is based on a human software engineer's daily practice. When a human developer tries to fix a bug, they would examine various resources such as the issue descriptions, error stack, debugging information, and test cases, until a solution is identified. However, currently most LLM-based repair approaches underutilize or use limited software artifacts. Here, we provide three examples to demonstrate the motivation of this work and the ability of LLMs to localize and fix bugs when given access to different types of software artifacts.

\subsection{Debugging Information to the Rescue}
\begin{figure}[htb] 
    \includegraphics[width=0.8\linewidth]{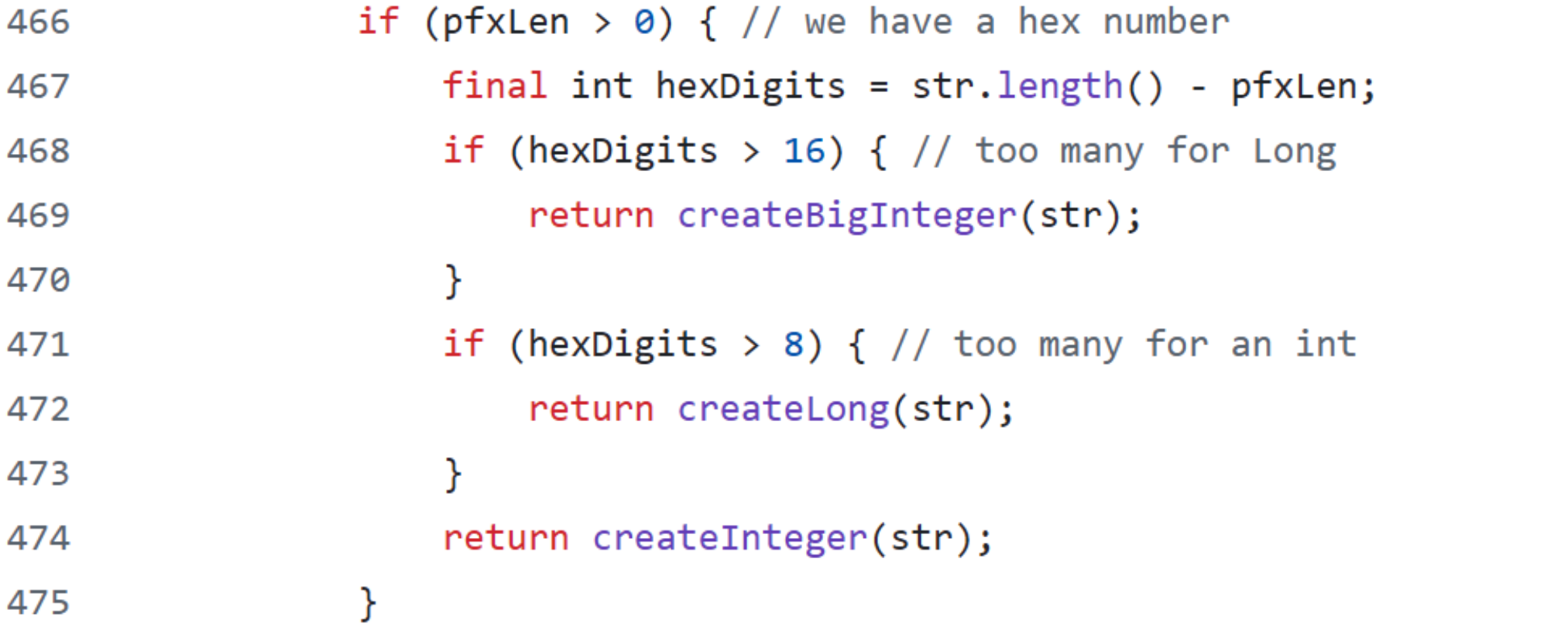}
    \caption{Code snippet for Lang 1b}
    \label{fig:lang1b_code_snippet}
\end{figure}

The Lang 1b bug involves the \textit{createNumber} method, which takes a string as its parameter and returns a number represented by that string. While studying the Lang 1b bug, we noticed that the newly added failing test case is the input ``0x80000000'' for the \textit{createNumber} method. When this test case is provided, the buggy \textit{createNumber} method incorrectly treats the input string as an integer, when it should be a long integer instead. The code snippet in Figure~\ref{fig:lang1b_code_snippet} shows a portion of the buggy \textit{createNumber} method in Lang 1b. The local variable \textit{hexDigits} represents the valid hexadecimal digits. As human software developers, we can quickly recognize that at Line 471, the condition should be ``\textit{hexDigits > 8 || (hexDigits == 8 \&\& firstSigDigit > `7')}'' (\textit{firstSigDigit} refers to the first valid hexadecimal digit). 


\begin{figure*}
\centering
    \includegraphics[width=0.8\linewidth]{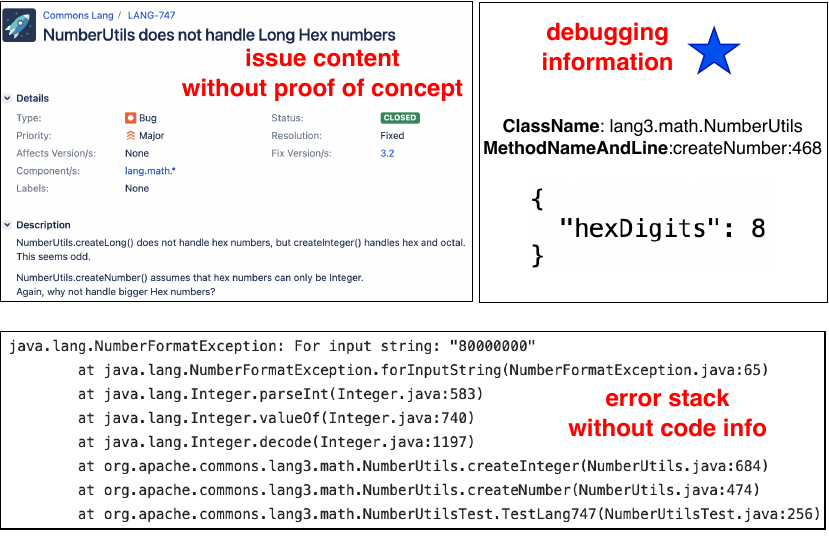}
    \caption{Issue content, error stack and debug info of Lang 1b}
    \label{fig:lang_1b}    
\end{figure*}

For this bug, we provided the LLM with Lang 1b's issue content (\url{https://issues.apache.org/jira/browse/LANG-747}) and the stack error message shown in Figure~\ref{fig:lang_1b}, but both could not help the LLM generate a plausible fix or near-to-correct patch. This is because the leap required in reasoning is significant, and the buggy method is quite long. Neither of the information provided above could pinpoint the buggy line precisely, let alone generate a plausible fix. To address this, we introduced debugging information to help the LLM understand the necessary changes. One piece of debug information we provided was a series of variable-value pairs, \textit{commons.lang3.math.NumberUtils:createNumber:468 \{hexDigits:8\}}. This indicates that the code is about to execute Line 468 and that the value of \textit{hexDigits} is 8 for the given test case. This enables the LLM to understand why the code on Line 474 was executed instead of Line 471. In one of the responses, the LLM suggested changing Line 468 to \textit{if (hexDigits > 16 || (hexDigits == 8 \&\& str.charAt(2) >= `8')) {}}, which, although not completely accurate, demonstrates that the provided debug information has a positive impact on fixing the bug.

\subsection{Issue Content to the Rescue}

\begin{figure}
    \centering
    \includegraphics[width=0.8\linewidth]{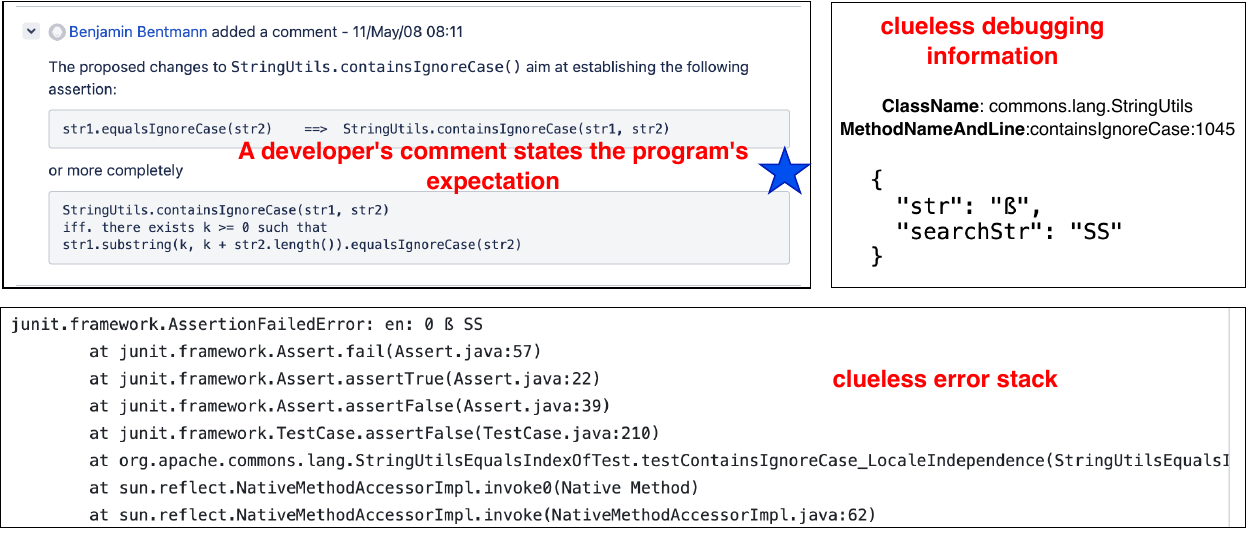}
    \caption{Issue content, error stack and debug info for Lang 40b}
    \label{fig:lang_40b}
\end{figure}

Sometimes, the discussions and Proof of Concept (PoC) included in the issue content can help the LLM better understand the expectations for bug repair. Figure~\ref{fig:lang_40b} shows a developer’s comment in the issue content, which states the expected behavior of the program (\url{https://issues.apache.org/jira/browse/LANG-432}). The original code attempted to convert both input strings to uppercase using \texttt{String.toUpperCase()} and then return the result of invoking the \textit{contains} method. However, \texttt{String.toUpperCase()} is locale-sensitive, which makes it unsuitable for case-insensitive comparisons. For example, the character \textbf{0x00DF} represents ``ß'' in Unicode. If we apply \texttt{String.toUpperCase()} to this character, it becomes ``SS'' in the Turkey locale. Consequently, comparing ``ss'' with ``ß'' would result in an equal match. Therefore, in this case, we should not use \texttt{String.toUpperCase()} but instead compare the characters individually.

The debug information, \textit{org.apache.commons.lang.StringUtils:containsIgnoreCase:1045} \textit{\{\leftquote{}str\rightquote{}:\leftquote{}ß\rightquote{},} \textit{\leftquote{}searchStr\rightquote{}:\leftquote{}SS\rightquote{}\}}, and the error stack trace shown in Figure~\ref{fig:lang_40b} merely re-display the test cases and do not clearly highlight the relationship between the character \textbf{0x00DF} and `SS'. As a result, neither the error stack trace nor the debug information provides any useful clues about the bug, and LLMs cannot generate a plausible fix when provided with either the stack message or the debug information alone. However, with the description and expected behavior outlined in the issue content, GPT-4o is able to generate a correct patch.

\begin{figure}
    \centering
    \includegraphics[width=0.8\linewidth]{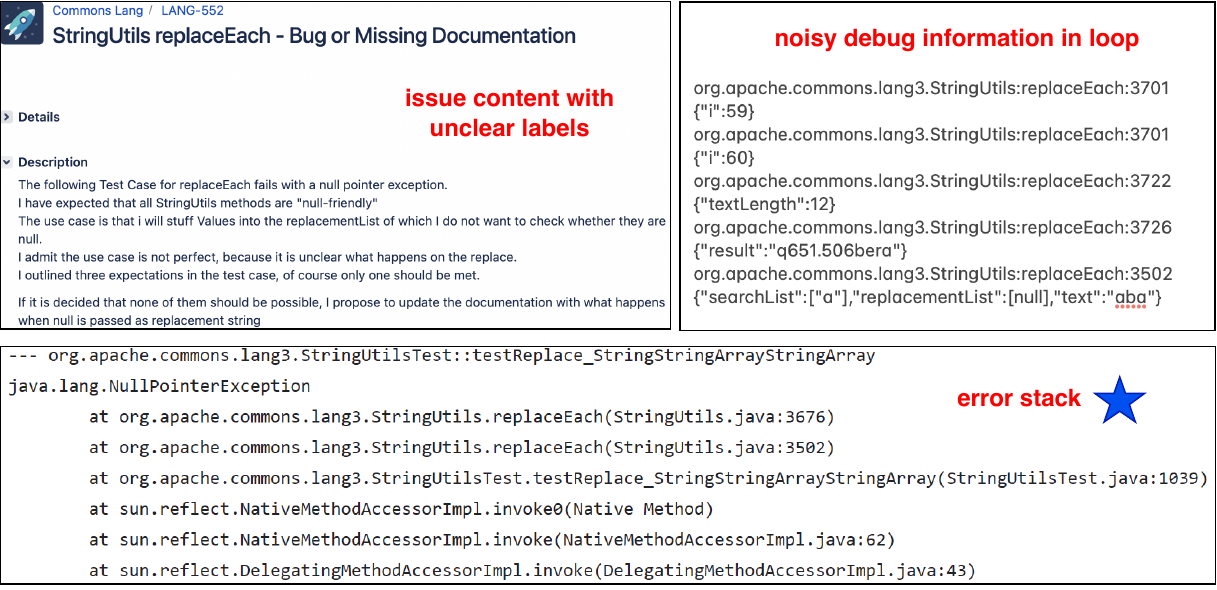}
    \caption{Issue content, error stack and debug info for Lang 39b}
    \label{fig:lang_39b_stack}
\end{figure}

\begin{figure}
    \centering
    \includegraphics[width=0.9\linewidth]{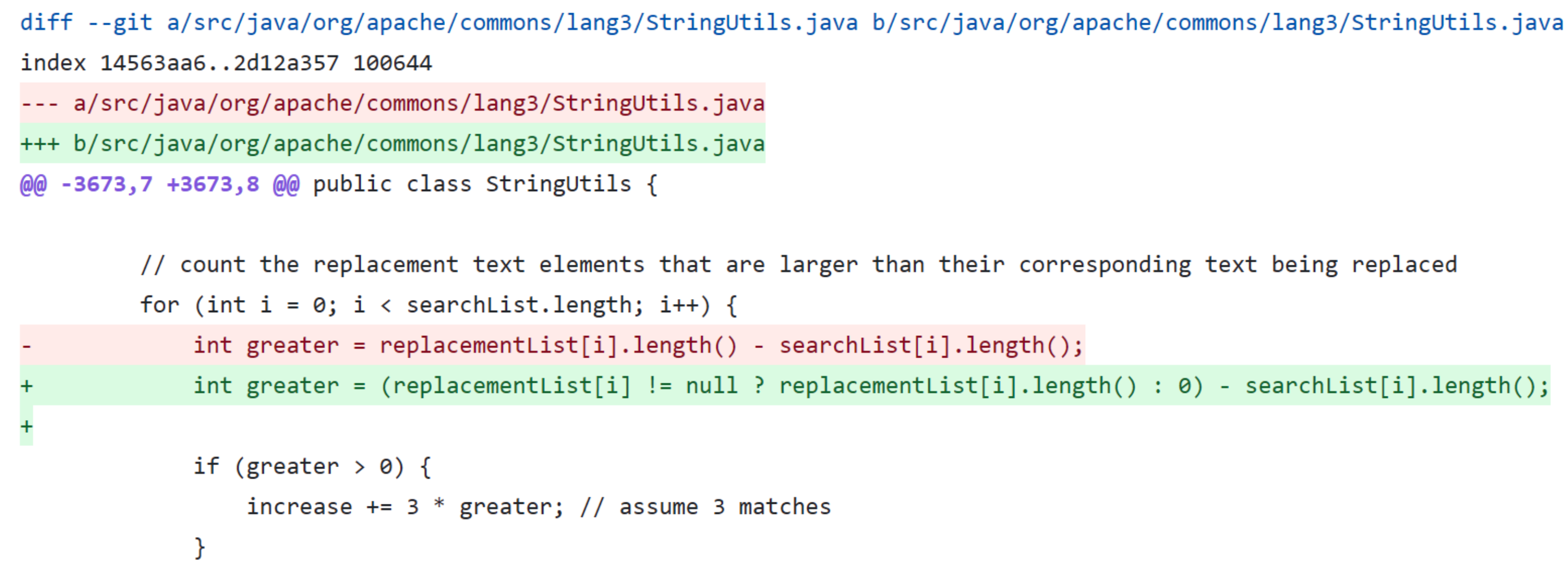}
    \caption{A patch generated by GPT-4o-mini for Lang 39b}
    \label{fig:lang_39b_llm_patch}
\end{figure}

\subsection{Error Stack to the Rescue}

When a bug triggers an exception, as opposed to merely being an unexpected behavior, the error stack trace generated at the point where the exception occurs can be helpful in diagnosing and fixing the error. Stack trace provides a detailed record of the sequence of method calls that led to an exception, making it easier to pinpoint the exact location in the code where the bug arises.

For example, in Lang 39b shown in Figure~\ref{fig:lang_39b_stack}, the developer forgot to check for \textit{null} values in the elements of the input parameters \textit{replacementList} and \textit{searchList}, which can lead to a NullPointerException (NPE). However, if we do not inform the LLM about the occurrence of an NPE in the error stack, the LLM will not be able to identify where \textit{null} detection is necessary. The stack trace clearly indicates that the NPE occurs on Line 3676 in the class \textit{StringUtils}. By providing this stack information to the LLM, it is able to generate a repair, as shown in Figure~\ref{fig:lang_39b_llm_patch}.

\subsection{Observations From the Above Three Examples}
From these three examples, we can see that issue content, error stack traces, and debugging information can complement each other in bug detection and fixing. Each type of artifact provides a different level of context, and when combined, they create a more comprehensive understanding of the bug. Issue content offers the high-level description and expectations for the fix, error stack traces pinpoint the exact location of the failure, and debugging information provides granular details about the variable states and flow of execution. By incorporating these diverse artifacts, the LLM’s repair process becomes more holistic.
\section{Approach}
\label{sec:approach}

\begin{figure}[h!]
    \centering
    \includegraphics[width=\linewidth]{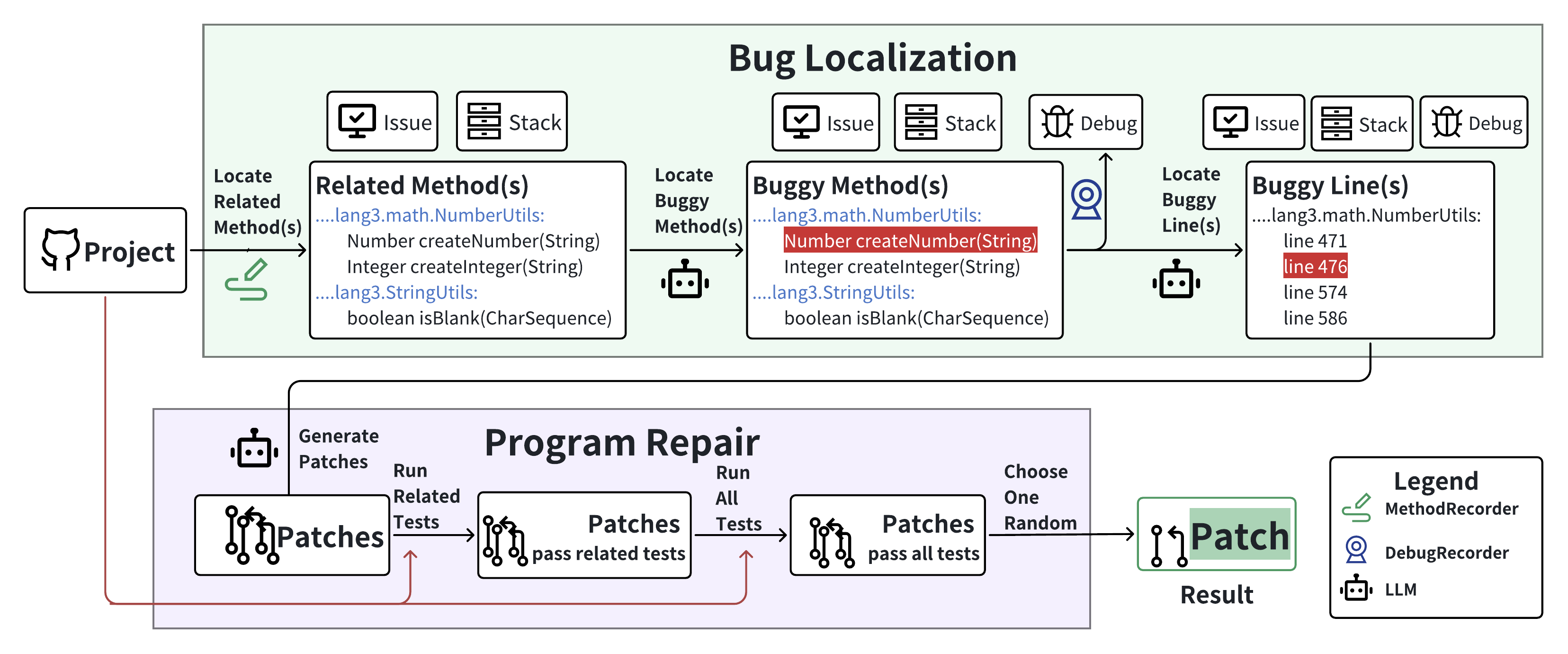}
    \caption{Our approach for bug localization and program repair}
    \label{fig:framework}
\end{figure}

As shown in Figure~\ref{fig:framework}, our approach utilizes LLMs in two distinct steps: bug localization and program repair. Along with these two steps, we developed two dynamic tools — \texttt{MethodRecorder} and \texttt{DebugRecorder} — to assist LLMs in handling bug localization and fix. \texttt{MethodRecorder} tracks the methods executed during the failing test case(s), while \texttt{DebugRecorder} is designed to extract debug information from the buggy method(s). Debug information is stored in a list, where each entry includes the method name, line number, variable names, and the JSON-formatted content of these variables. Table~\ref{tbl:prompt} shows the prompts used in our process. The first, second, and third rows correspond to the prompts used for the tasks of localizing buggy methods, localizing buggy lines, and generating patches, respectively. The first two rows of prompts are used for the bug localization task and the last row of prompt is used for the program repair task. For each task, we provide the LLM with <\textsc{General Task Prompt}> + <\textsc{Input Prompt}> + <\textsc{Expected Output Prompt}> to receive the response from the LLM.

\begin{table}
\caption{Prompts used in localizing buggy methods, buggy lines and generating patches}
    \label{tbl:prompt}
    \centering
\begin{tabular}{>{\hspace{0pt}}m{0.26\linewidth}>{\hspace{0pt}}m{0.2\linewidth}>{\hspace{0pt}}m{0.5\linewidth}} \hline
\textsc{General Task Prompt}  & \textsc{Input Prompt}  & \textsc{Expected Output Prompt} \\\hline
You are a Software Engineer. Review the following skeleton of classes, test case(s), and exception that occurs when doing the test. 
\par{}Provide a set of locations that need to be edited to fix the bug. \textbf{The locations must be specified as method names or field names.}
& \textbf{\#\#\# Skeleton of Classes \#\#\#} 
\par{}  \textcolor{magenta}{\{related\_methods\}}
\par{} \textcolor{black}{\{error\_stack\}} 
\par{} \textcolor{teal}{\{issue\_content\}}
& Please \textbf{localize class name and method names or field names} that need to be edited. 
\par{}Examples:
\par{} path.to.ClassA::methodA
\par{} path.to.ClassA::methodB
\par{} path.to.ClassB::methodA  \\ \hline

You are a Software Engineer. Review the following skeleton of classes, test case(s), and exception that occurs when doing the test. \par{} Provide a set of locations that need to be edited to fix the bug. \textbf{The locations must be specified as line number in class.}
& \textbf{\#\#\# Skeleton of Classes \#\#\#} 
\par{} \textcolor{red}{\{buggy\_methods\}}
\par{} \textcolor{black}{\{error\_stack\}}
\par{} \textcolor{teal}{\{issue\_content\}}
\par{} \textcolor{cyan}{\{debugging\_info\}} 

& Please \textbf{localize class name and line number} that need to be edited. 
\par{}Examples:
\par{} path.to.ClassA
\par{} ~~line: 20
\par{} ~~line: 45
\par{} ~~line: 46
\par{} ~~line: 47  \\ \hline

You are a Software Engineer. Review the following methods and(or) fields of classes, test case(s), and exception that occurs when doing the test. \textbf{Try to fix the bug.}
&\textbf{\#\#\# Skeleton of Classes \#\#\#}\par{}\textcolor{red}{\{buggy\_methods\}}
\textbf{\#\#\# Possible bug locations (for your reference only) \#\#\#}
\par{}\textcolor{orange}{\{buggy\_lines\}}
\par{} \textcolor{black}{\{error\_stack\}}
\par{} \textcolor{teal}{\{issue\_content\}}
\par{} \textcolor{cyan}{\{debugging\_info\}}

& Please generate *SEARCH/REPLACE* edits to \textbf{fix the bug} based on the info given above. Every *SEARCH/REPLACE* edit must use this format:
\par{} ~~1. The file path 
\par{} ~~2. The start of search block: \textless{}\textless{}\textless{}\textless{}\textless{}\textless{}\textless{} SEARCH
\par{} ~~3. A contiguous chunk of lines to search in the existing source code
\par{} ~~4. The dividing line: =======
\par{} ~~5. The lines to replace into the source code
\par{} ~~6. The end of the replace block: \textgreater{}\textgreater{}\textgreater{}\textgreater{}\textgreater{}\textgreater{}\textgreater{} REPLACE  \\ \hline
\end{tabular}
\end{table}

\subsection{Localize Buggy Method(s)}
We emulate the behavior of human developers, who use errors in stack traces and proof of concept (PoC) from issue content to localize buggy methods. To this end, we apply \texttt{MethodRecorder} to trace the methods invoked during the execution of failing test cases. We then collect the signatures of these invoked methods and prompt the LLM to identify the buggy method(s). In this step, we also provide the LLM with any available error stack traces and issue content, including general description of the issue and developers' discussion/comment under this issue. We choose not to use debugging information because it is typically gathered at runtime and involves instrumentation (see \textit{java.lang.instrument.Instrumentation}~\cite{java8instrumentation}), which can be computationally expensive. In developers' daily practice, they set just a few of breakpoints and collect important information about variables. But setting breakpoints automatically is not practical. Therefore, before localizing buggy method(s), minimizing instrumentation and debugging information is more efficient.

\subsection{Localize Buggy Line(s)}
Once the buggy method(s) have been localized in the previous step, we utilize \texttt{DebugRecorder} to capture detailed information about variable names and their values within the identified buggy method(s). The rationale for introducing debugging information at this stage is that it provides a more granular view, which is particularly beneficial for pinpointing the specific lines responsible for the bug and for understanding the internal state of variables. The debugging information with variables' value is often unnecessary during the initial localization step, but becomes crucial when the focus shifts to identifying precise faults within a known problematic method.

In this process, we gather debugging information as a list of variable-value pairs that reflect the state at specific lines of code within the buggy method(s). This allows the LLM to understand the conditions and data flows that may contribute to the bug. Combined with the issue content, such as descriptions and discussions, along with stack traces, this debugging information is fed into the LLM along with the body of the buggy method(s). Together, these resources assist the LLM in localizing the exact line(s) within the method(s) where the bug manifests, providing the context needed to understand and resolve the bug effectively.

\subsection{Patch Generation}
For patch generation, we provide all relevant information to maximize the LLM’s ability to produce an accurate repair. This includes the localized buggy buggy line(s) and method(s) identified in previous steps, as well as the complete body of the buggy method(s). Additionally, we provide supplementary materials such as the issue content, which may contain descriptions and discussions about the bug, the stack trace from the error, and detailed debugging information captured at runtime. By supplying this comprehensive context, we enable the LLM to better understand the specific code behavior that led to the error and to use this knowledge to generate a more targeted and effective patch for the bug. For each bug, we use the LLM to generate multiple potential patches in \textit{.diff} format, which can be directly applied to the buggy code.

\subsection{Patch Validation}
\label{patch_validation}

We apply the patches generated in the previous step to the buggy code and test them to evaluate their effectiveness. If the patch passes the initial failed tests, we proceed it to pass all tests in case that it incur any regression. If it passed all the tests, then it is considered a plausible fix. To facilitate this process, we utilize the Defects4J framework to run the tests and verify the patches. 

To be consistent with existing studies~\cite{zhu2023tare,li_giantrepair_2024}, if the patch is semantically equivalent to the original patch provided by developers, it is considered a correct patch. As part of this validation process, the patch undergoes manual inspection and cross-checking by two experienced developers to ensure its correctness. This step is crucial to verify that the patch not only passes all automated tests but also aligns with the intended behavior from a developer’s perspective. The manual inspection by developers serves as a final quality control step to ensure that the patch addresses the bug correctly and does not introduce new bugs.
\section{Experiment Setup}
\label{sec:experiment}

\subsection{Dataset}
We used the well-established Defects4J benchmark~\cite{defects4j} for our experiments, specifically leveraging both version 1.2 and version 2.0. Defects4J v1.2 contains 391 bugs from six real-world projects (note that there are 395 bugs in v1.2, but due to the update to Java 8, four bugs can no longer be reproduced), while v2.0 includes an additional 444 bugs from 11 real-world projects. We chose Defects4J benchmark for two main reasons. First, it offers a diverse set of software artifacts — such as issue URLs, error stacks, and test cases — that are well-suited for our proposed ARP framework and help maximize its capabilities. Second, Defects4J benchmark has been widely used in prior research, which facilitates a fair and meaningful comparison of our approach. 

\subsection{Parameters in Experiments and Baseline Selection}

\subsubsection{Parameters in Experiments:} Due to the extremely long debugging information for some bugs recorded by \texttt{DebugRecorder}, along with issue content and error stack, which exceeds the token limit of most LLMs, we opted for the GPT-4 series model, known for its ability to handle long token sequences with a 128k token context window. To manage costs, we selected the cost-effective \textbf{GPT-4o-mini} model, also with a 128k token context window, and used the default settings (\textit{Temperature}=0.5, Top-\textit{p}=1). \textcolor{black}{For the experimental setup, before automating the entire process, we manually used a client with access to multiple LLMs and provided artifact information for 10 random selected Defects4J bugs to evaluate the feasibility of our approach and to select appropriate models—considering factors such as low cost, a large token context window, and minimal hallucination rate. During this step, we found that the GPT-3.5 series exhibited a high rate of hallucination when generating patches, and we excluded it from consideration. We also determined the appropriate format for the artifacts to be fed into the LLM. For example, we pruned the debug information such as only recording variables that have changed, cropping long arrays and lists to minimize token size. Additionally, we instructed the LLM to generate *SEARCH/REPLACE* edits when producing patches, in order to reduce hallucination rate, as adopted by SOTA approaches~\cite{xia_agentless_2024,ouyang2024repograph,zhang2024autocoderover,liu2024marscode}. After selecting the model, prompts and artifacts format, we used this preliminary process to determine the optimal number of LLM calls at each step: localizing buggy methods, localizing buggy lines, and generating patches. The rationale is that increasing the number of LLM calls beyond a certain point does not lead to significant performance gains.} Based on these manual experiments, we finalized the settings of the parameters. For buggy method localization, we call the LLM only once. For buggy line localization, we call the LLM 10 times, collecting 10 responses, each identifying potential buggy lines. We then filter out duplicate responses (i.e., identical buggy line locations). Finally, for each unique response, we ask the LLM to generate patches three times, ensuring that multiple potential fixes are considered.

{\color{black} \subsubsection{Baseline Selections:} We selected state-of-the-art LLM-based fault localization and program repair frameworks on the Defects4J dataset as our baselines for two reasons. First, using the same Defects4J dataset allows us to directly compare our results with those reported in the baseline papers. Second, we focus on LLM-based approaches to better evaluate our framework, which is also LLM-based and leveraging a set of software artifacts.}

\subsection{Research Questions}
In this work, we aim to answer the following research questions (RQs) for evaluating the effectiveness of various software artifacts in assisting the chosen LLM in bug localization and APR task.

\begin{itemize}
    \item \textbf{RQ1. Which software artifacts can better assist LLMs in localizing buggy methods and buggy lines when provided with source code?} This RQ aims to identify the specific types of software artifacts - issue content, debugging information and error stack trace - that enhance the ability of LLMs to accurately pinpoint methods and code that contain bugs. 
    
    \item \textbf{RQ2. Which software artifacts can better assist LLMs in generating plausible patches when provided with buggy methods?} By answering this RQ, we can identify the specific types of software artifacts in assisting LLMs to produce plausible patches which can pass all unit tests.
    
    \item \textbf{RQ3. What is the overall performance of \toolName{} in bug localization and program repair?} Unlike some LLM-based approaches that focus on either fault localization or program repair while relying on traditional methods such as spectrum-based approaches~\cite{ochiai} for fault localization~\cite{li_giantrepair_2024,fitrepair}, \toolName{} supports an end-to-end bug identification and program repair workflow. This RQ aims to evaluate the overall performance of \toolName{} from the initial input of a code repository to the final output of a complete repair.

\end{itemize}

\section{Results}
\label{sec:results}

\subsection{RQ1. Buggy Method and Buggy Lines Localization from Code Repository}
\label{rq1}

\subsubsection{Localizing Buggy Method(s)} We first use our \texttt{MethodRecorder} tool, which runs the failing tests and records the signatures of the methods that have been executed to narrow down the scope of buggy methods. Then we provide the LLM with the signatures of the executed methods, along with issue content (including the issue description and discussion, which we crawled from each bug's issue URL) and the error stack trace corresponding to the failing test case, as supplied by the Defects4J framework. In this step, we choose not to generate the debugging information. The main reason is that the debugging information can be too long for LLMs to process effectively before the buggy method(s) have been localized. Using the prompt specified in Table~\ref{tbl:prompt}, we ask the LLM to output the signatures of buggy method(s). To evaluate what kinds of software artifacts can better assist the LLM in localizing buggy methods, we conduct four separate experiments by feeding the LLM with (1) only executed method signatures, (2) executed method signatures and issue content, (3) executed method signatures and error stack trace, and (4) executed method signatures, issue content and error stack trace.

\begin{table}[!ht]
    \begin{minipage}{0.5\linewidth}
    \caption{Effectiveness of buggy method localization in Defects4J v2.0 with different software artifacts}
    \label{tbl:rq1-method-localization}
    \centering
    \begin{tabular}{c|ll}    
    \hline
        ~   & \textbf{Single Method}  & \textbf{Non-Single} \\ \hline
        --- & 135/486=27.8\%  & 77/320=24.1\%\\ 
        Issue    & 207/475=43.6\%  & 127/313=40.6\%\\
        Stack   & 216/486=44.4\%  & 123/320=38.4\%\\ 
        Issue + Stack & 234/475=49.3\%  & 149/313=47.6\%\\ \hline
    \end{tabular}
    \end{minipage}\hfill
	\begin{minipage}{0.45\linewidth}
		\centering
		\includegraphics[width=\linewidth,trim=200 10 0 190,clip]{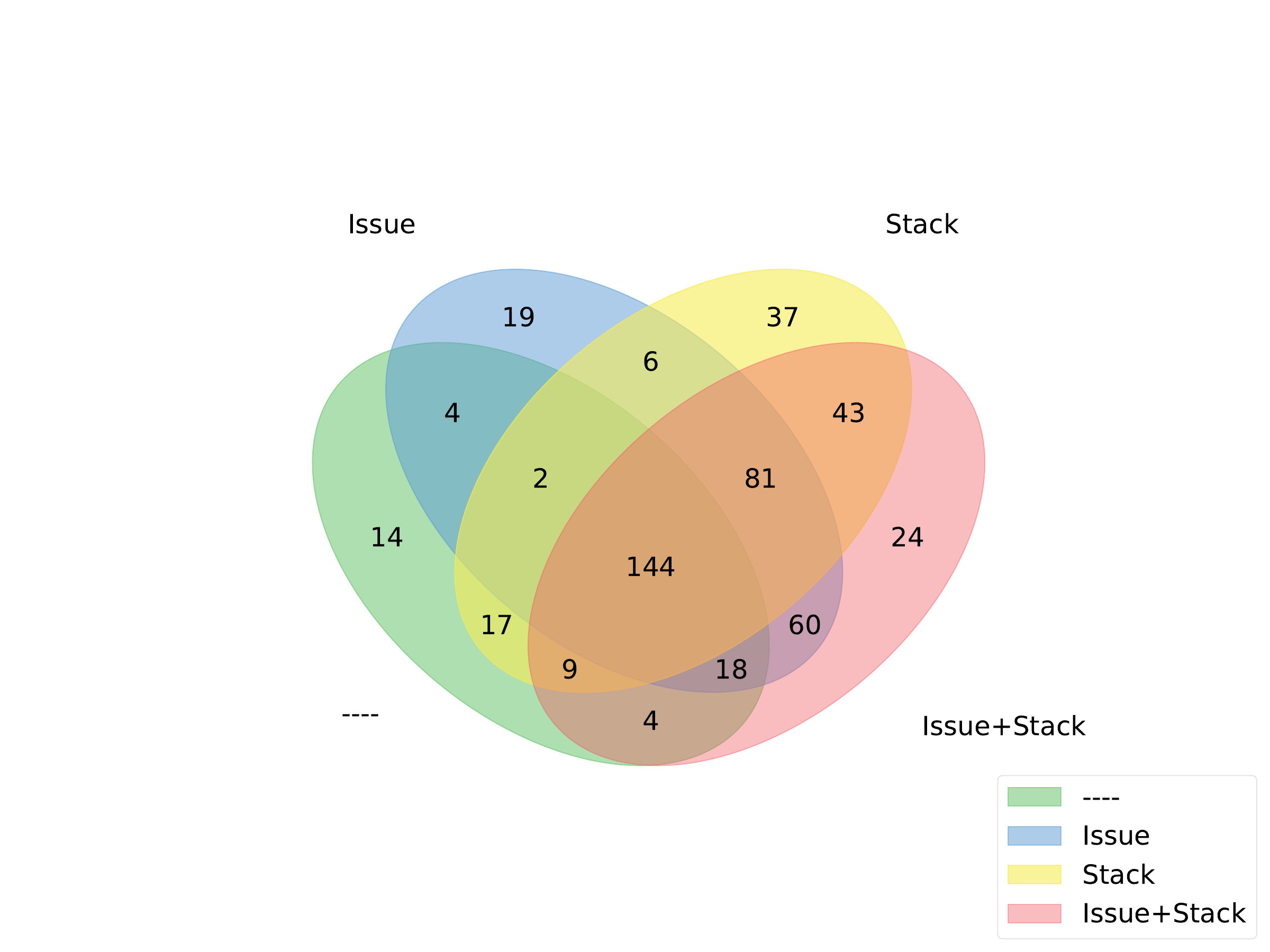}
		\captionof{figure}{Overlap of buggy method localization in Defects4J v2.0}
    \label{fig:buggy_method_localization_overlap}
	\end{minipage}
\end{table}

Table~\ref{tbl:rq1-method-localization} shows the effectiveness of buggy method localization in these four experiments. The first row, labeled \leftquote{}---\rightquote{}, indicates that only the related (executed) methods are fed to the LLM for buggy method localization. The second, third and fourth rows represent scenarios when, in addition to related method localization, issue content, error stack, or both are provided. Following the approach described in LLMAO~\cite{yang_large_2024}, we consider a buggy method correctly localized if at least one match in the LLM's response corresponds to the buggy method (single-method bugs) or one of the buggy methods (non-single-method bugs) in the ground truth. It is worth mentioning that there exist bug cases where certain software artifacts can not be extracted, and in this case we calculate the ratio by dividing only the number of bug cases with certain software artifacts information. Table~\ref{tbl:rq1-method-localization} shows that, when issue content or error stack is provided, the LLM is able to localize more buggy methods than nothing is provided. When issue content is provided, 15.8\% more single buggy methods and 16.5\% more non-single buggy methods are correctly localized. When error stack is included, 16.6\% more single buggy methods and 14.3\% more non-single buggy methods are correctly localized. Furthermore, combining issue content and error stack achieves the best performance, which results in 49.3\% in localizing single buggy methods and 47.6\% of non-single buggy methods.

\begin{table}[]
    \centering
    \footnotesize
   \caption{\textcolor{black}{Method-level FL comparison with SOTA baselines on Defects4J v1.2}}
    \label{tbl:compare_FL_sota}
   {\color{black}\begin{tabular}{cc|ccc|ccc|ccc|ccc}\hline
         \multirow{3}{*}{Project} & \multirow{3}{*}{\#Bugs} & \multicolumn{3}{c|}{\toolName{}*} & \multicolumn{3}{c|}{SoapFL~\cite{soapfl}} & \multicolumn{3}{c|}{AutoFL~\cite{autofl}} & \multicolumn{3}{c}{Agentless~\cite{xia_agentless_2024}} \\ \cline{3-14}
         ~ & ~ & \multicolumn{3}{p{2cm}|}{codebase, issue content, failed tests, error stack} & \multicolumn{3}{p{2cm}|}{codebase, failed tests, error stack} & \multicolumn{3}{p{2cm}|}{failed tests, error stack} & \multicolumn{3}{p{2cm}}{codebase, issue message} \\ \cline{3-14}
         ~ & ~ & Top1 & Top3 & Top5 &Top1&Top3&Top5& Top1 & Top3 & Top5 & Top1 & Top3 & Top5 \\ \hline
         Chart &  26 &  18 &  18 &  18 &  17 &  22 &  22 &  16 &  20 &  21 &  10 &  15 &  16  \\
Lang &  65 &  54 &  58 &  58 &  39 &  48 &  49 &  35 &  43 &  43 &  34 &  40 &  48  \\
Math &  106 &  81 &  86 &  86 &  57 &  71 &  73 &  53 &  67 &  70 &  45 &  51 &  52  \\
Time &  27 &  12 &  13 &  13 &  10 &  13 &  13 &  14 &  15 &  16 &  8 &  10 &  12  \\
Mockito &  38 &  22 &  22 &  22 &  20 &  21 &  22 &  13 &  17 &  17 &  21 &  26 &  26 \\
Closure &  133 &  26 &  37 &  41 &  32 &  57 &  57 &  18 &  24 &  24 &  18 &  31 &  31 \\ \hline
Overall &  395 &  213 &  234 &  238 &  175 &  232 &  236 &  149 &  186 &  191 &  136 &  173 &  185  
\\ \hline
    \end{tabular}}
    \\\color{black}{*: To ensure a fair comparison, \toolName{} uses the same GPT-3.5 model as the other baselines.}
\end{table}

Figure~\ref{fig:buggy_method_localization_overlap} shows the overlap of results of buggy method localization with different artifacts in Defects4J v2.0. We can see that when provided with issue content, the LLM can localize 19 extra bugs correctly. Similarly, when provided with error stack, the LLM can localize 37 extra bugs correctly. This finding verifies our assumption in Section~\ref{sec:motivation} that different kinds of software artifacts can complement each other in buggy method localization.

As existing SOTA fault localization approaches use Defects4J v1.2 dataset and adopt Top-\textit{n} (the buggy method is in the top $n$ list of the LLM's response) to evaluate the performance of fault localization, we compare the performance of \toolName{} with the SOTA approaches in the same dataset and adopt Top-\textit{n} approach. \textcolor{black}{Table~\ref{tbl:compare_FL_sota} presents the comparison results of \toolName{} with three state-of-the-art (SOTA) LLM-based approaches. The row below the names of these approaches lists the artifacts used by each approach. Among them, \texttt{SoapFL}~\cite{soapfl} uses codebase, failed test methods, and error stack trace; \texttt{AutoFL}~\cite{autofl} uses failed test methods along with error stack trace; and \texttt{Agentless}~\cite{xia_agentless_2024} uses codebase and issue message. In contrast to these approaches, \toolName{} utilizes four artifacts: issue content, error stack, failed tests, and codebase. Failed tests and codebase are used to retrieve related methods via \texttt{MethodRecorder}. It is worth to mention that \toolName{} also outperform the SOTA learning-based approaches such as \texttt{GRACE}~\cite{lou2021boosting} and \texttt{FLUCCS}~\cite{sohn2017fluccs}. But in this study we focused on how various artifacts can help LLMs with bug localization and program repair, so we did not include the comparison with non-LLM-based approaches. Since the other three baselines use the GPT-3.5 model, for a fair comparison, \toolName{} also uses the same GPT-3.5 model. As we can see, \toolName{} outperforms \texttt{SoapFL}, \texttt{AutoFL} and \texttt{Agentless} in Top-1, Top-3 and Top-5 metrics. The last row shows that, \toolName{} outperforms other approaches in Top-1 and achieves similar performance in Top-3 and Top-5 with \texttt{SoapFL}, demonstrating its ability to precisely localize buggy methods.}

\begin{table}[]
    \centering
    \caption{Effectiveness of buggy line localization with different software artifacts}
    \label{tab:line_precision}
    \begin{tabular}{c|ccc}
    \hline
    ~ & \textbf{Match} & \textbf{Single Method} & \textbf{Non-Single} \\ \hline
\multirow{3}{*}{---} & Exact & 40.3\% & 4.3\%\\ 
		~& Range-3 & 68.1\% & 8.1\% \\ 
		~& Range-5 & 72.8\% & 9.8\% \\
\multirow{3}{*}{Issue} & Exact & 49.5\% & 5.2\% \\ 
		~& Range-3 & 74.4\% & 9.2\% \\ 
		~& Range-5 & 77.5\% & 11.0\% \\
\multirow{3}{*}{Stack} & Exact & 45.0\% & 4.9\% \\ 
		~& Range-3 & 66.9\% & 9.0\% \\ 
		~& Range-5 & 71.0\% & 10.1\% \\
\multirow{3}{*}{Debug } & Exact & 39.9\% & 3.5\% \\ 
		~& Range-3 & 65.4\% & 6.4\% \\ 
		~& Range-5 & 70.3\% & 8.1\% \\
\multirow{3}{*}{Stack + Debug} & Exact & 44.8\% & 5.2\% \\ 
		~& Range-3 & 66.9\% & 9.0\% \\ 
		~& Range-5 & 70.3\% & 10.7\% \\
\multirow{3}{*}{Issue + Debug} & Exact & 50.1\% & 5.2\% \\ 
		~& Range-3 & 70.3\% & 8.4\% \\ 
		~& Range-5 & 73.2\% & 9.8\% \\
\multirow{3}{*}{Issue + Stack} & Exact & 49.1\% & 5.5\% \\ 
		~& Range-3 & 73.2\% & 10.1\% \\ 
		~& Range-5 & 77.5\% & 11.8\% \\
\multirow{3}{*}{Issue + Stack + Debug} & Exact & 50.3\% & 5.2\% \\ 
		~& Range-3 & 71.4\% & 11.3\% \\ 
		~& Range-5 & 73.8\% & 12.7\% \\
\multirow{3}{*}{Union} & Exact & 68.9\% & 9.0\% \\ 
		~& Range-3 & 83.8\% & 16.5\% \\ 
		~& Range-5 & 85.7\% & 19.1\% \\   \hline
    \end{tabular}
\end{table}

\subsubsection{Localizing Buggy Line(s)}
To evaluate LLMs’ ability in localizing buggy lines, we provide the LLM with the buggy method(s) baseline (i.e., the ground truth which can be extracted from human developer's correct patch) and ask the LLM to predict which lines are buggy. If the line number generated by the LLM exactly matches the first line added in the human developer's correct patch in Defects4J, we consider it an exact match. If the predicted line number falls within a range of $n$ lines from the first added line in the human developer's patch (for example, if the correct patch starts at Line 368, and the LLM's prediction falls within the range from $368-n$ to $368+n$), we classify it as a Range-\textit{n} match.

\begin{figure}[!tbp]
    \centering
    \includegraphics[width=0.6\textwidth]{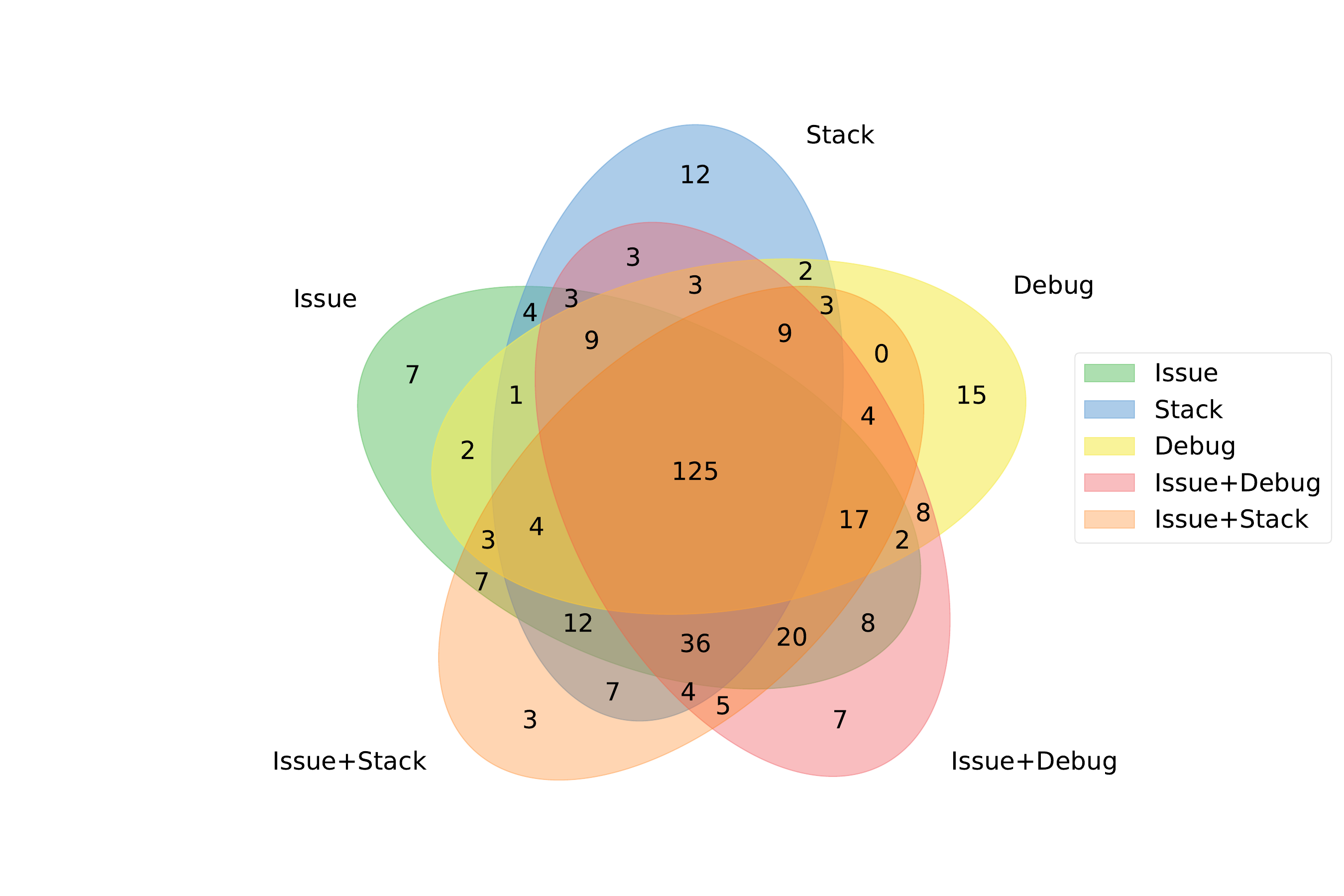}
    \caption{Buggy line localization exact match with different artifacts in Defects4J v2.0}
    \label{fig:buggy_line_localization_overlap}
\end{figure}

Table~\ref{tab:line_precision} shows the effectiveness of buggy line localization with different artifacts. We can see that when the LLM is fed with issue content, it performs the best among all single software artifacts, achieving 49.5\% in exact matches, 74.4\% in Range-3, and 77.5\% in Range-5 for single methods. \textcolor{black}{When adding more software artifacts, the performance does not appear to improve. For example, combining issue content with the error stack does not yield better exact, Range-3, or Range-5 buggy line matches for single methods compared to using issue content alone. However, as seen from Figure~\ref{fig:buggy_line_localization_overlap}, the combination does successfully localize 3 additional buggy lines that cannot be identified using either issue content or the error stack individually.} The last row indicates that, when combining all artifacts, 68.9\% of buggy lines in single-method bugs and 9\% of buggy lines in non-single-method bugs are correctly localized. Similar to Figure~\ref{fig:buggy_method_localization_overlap}, Figure~\ref{fig:buggy_line_localization_overlap} demonstrates that, while most of the correctly localized buggy lines across different artifacts overlap, different combinations of artifacts each have their own advantages (since only five areas can be displayed in the overlap figure, we randomly select five categories for visualization). For example, using only the issue content can correctly localize 7 more buggy lines, while using only the debug information can localize 15 more.

\noindent\begin{center}
		\begin{tcolorbox}[colback=black!5, colframe=black!20, width=1.0\linewidth, arc=1mm, auto outer arc, boxrule=1.5pt]                       
            {{\textbf{Answer to RQ1:} Among all three types of single software artifacts, issue content is the most effective in assisting the LLM with bug localization. Furthermore, different types of software artifacts complement each other in localizing buggy methods and buggy lines. This aligns with how human developers typically approach bug localization, as they often rely on information from multiple sources to identify bugs.}}
		\end{tcolorbox}
\end{center}
\subsection{RQ2. Program Repair Based on Provided Method-level Fault Localization}
\label{rq2}

\begin{table}[!ht]
    \begin{minipage}{0.6\linewidth}
    \caption{Plausible patch with provided method-level fault localization in Defects4J v2.0}
    \label{tbl:patch_precision}
    \centering
    \begin{tabular}{c|ll}
    \hline

        ~ & Single Method & Non Single \\ \hline
        --- &46/489=9.4\% &4/345=1.2\%\\
        Issue &127/478=26.6\% &14/337=4.2\%\\
        Stack &132/489=27.0\% &13/345=3.8\%\\
        Debug  &114/468=24.4\% &11/326=3.4\%\\
        Issue+Stack &159/464=34.3\% &21/328=6.4\%\\
        Issue+Debug &191/457=41.8\% &24/318=7.6\%\\
        Stack+Debug&139/462=30.1\%&14/321=4.4\%\\
        Issue+Stack+Debug &197/457=43.1\% &30/318=9.4\%\\
        Union &274/489=56.0\% &50/346=14.5\%\\ \hline
    \end{tabular}
    \end{minipage}\hfill
 \begin{minipage}{0.4\linewidth}
    \centering
    \includegraphics[width=\textwidth,trim=240 20 0 20,clip]{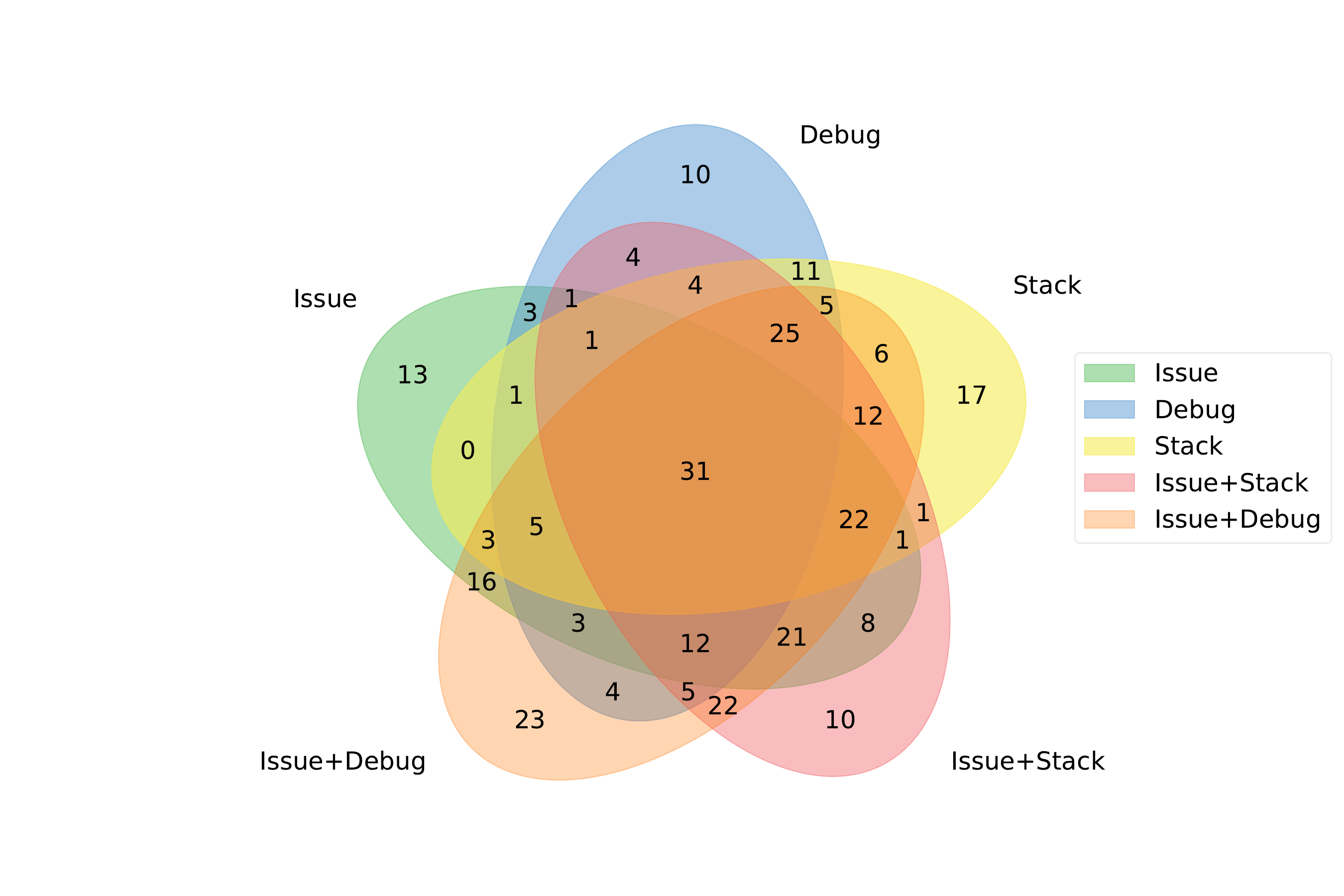}
    \captionof{figure}{Overlap of program repair with different artifacts in Defects4J v2.0}
    \label{fig:patch_overlap_artifacts}
\end{minipage}

\end{table}

To evaluate the effectiveness of program repair, we first extract buggy methods from human's correct patches and used these buggy methods as the baseline. Then we use \texttt{DebugRecorder} to get the debug information in these buggy methods. Finally, we provide the LLM with different artifacts and check how many plausible patches (passing all tests) it could generate. 

Table~\ref{tbl:patch_precision} shows that, error stack achieves the best performance in generating plausible patches for single-method bugs (27.0\%) among all three software artifacts, while issue content performs the best for non-single buggy methods. If feeding the LLM with all three types of software artifacts, we can get the best performance, which is 43.1\% and 9.4\% in patching single-method bugs and non-single-method bugs, respectively. Figure~\ref{fig:patch_overlap_artifacts} shows that different artifacts complement each other in program repair. Error stack can help the LLM fix 17 extra bugs while issue content can fix 13 extra bugs. Combining issue content with debug info can assist the LLM in fixing 23 extra bugs. 

\begin{figure}[!tbp]
    \centering
    \includegraphics[width=0.5\textwidth,trim=240 20 0 20,clip]{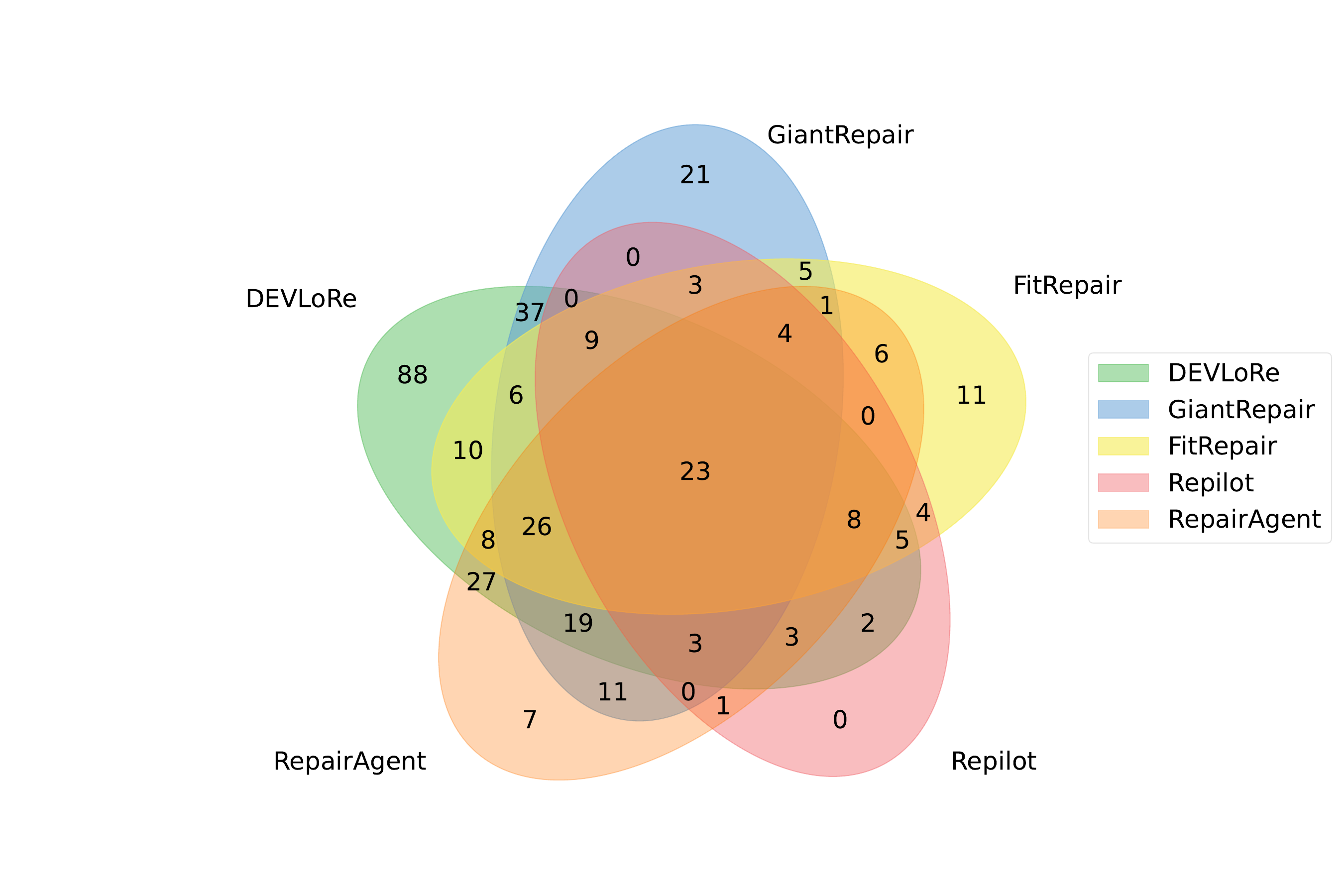}
    \caption{\textcolor{black}{A Comparison between \toolName{} with other LLM-based SOTA program repair approaches}}
    \label{fig:patch_overlap}
\end{figure}

\textcolor{black}{Table~\ref{tbl:compare_patch_sota} compares \toolName{} with other state-of-the-art (SOTA) LLM-based program repair approaches. Since these approaches focus on repairing single-method bugs, we evaluate our tool on the same dataset for a fair comparison. The row below the names of these approaches lists the artifacts used by each approach. Specifically, \texttt{GiantRepair}~\cite{li_giantrepair_2024} and \texttt{Repilot}~\cite{repilot} take the buggy function as input; \texttt{RepairAgent}~\cite{repairagent} uses the failed tests and all methods in the codebase; \texttt{ChatRepair}~\cite{chatrepair} uses the failed tests and the buggy line; and \texttt{FitRepair}~\cite{fitrepair} uses only the buggy lines within the buggy method. In contrast, \toolName{} leverages four artifacts including the buggy method, debugging information, stack traces, and issue content.} As shown in Table~\ref{tbl:compare_patch_sota}, 142 out of 259 bugs in Defects4J v1.2 and 132 out of 230 bugs in Defects4J v2.0 can be repaired by \toolName{}, outperforming the other five LLM-based approaches. In total, \toolName{} can fix 274 buggy methods, which is 60.2\% more than \texttt{GiantRepair}~\cite{li_giantrepair_2024}. \textcolor{black}{Furthermore, Figure~\ref{fig:patch_overlap} shows that 88 bugs can only be fixed by \toolName{}. This result demonstrates \toolName{}'s ability in fixing more and extra bugs by leveraging different types of software artifacts.}

\begin{table}[!ht]
    \footnotesize
    \caption{\textcolor{black}{Plausible patches with provided method-level fault localization in \textbf{single buggy method} projects}}
    \label{tbl:compare_patch_sota}
    \centering
    {\color{black}\begin{tabular}{ccp{2cm}p{1.2cm}p{1.5cm}p{1.5cm}p{1.5cm}p{1.2cm}} \hline
         \multirow{2}{*}{Project} & \multirow{2}{*}{\#Bugs}   &  \toolName{}  & GaintRepair & RepairAgent & ChatRepair & FitRepair & Repilot\\ \cline{3-8}
        ~ & ~ & buggy method, issue content, debugging info, stack traces  & buggy method &  failed tests, all methods in codebase &  failed tests, buggy lines & buggy lines within buggy method & buggy method \\ \hline
        Chart            & 16 & 12 & 8 & 11 & 15 & 8 & 6   \\
        Closure          & 95 & 37 & 32 & 27 & 37 & 29 & 21  \\
        Lang             & 42 & 32 & 14 & 17 & 21 & 17 & 15   \\
        Math             & 74 & 53 & 26 & 29 & 32 & 23 & 20   \\
        Time             & 16 & 3 & 1 & 2 & 6 & 3 & 2  \\
        Mockito          & 16 & 10 & 6 & 6 & 3 & 4 & 0  \\ \hline
Defects4J v1.2 & 259 & 142 & 87 & 74 & 114 & 85 & 64  \\
Defects4J v2.0 & 230 & 132 & 84 & 90 & 48 & 44 & 47  \\ \hline
        Total           & 489 & 274 & 171 & 164 & 162 & 129 & 111 \\ \hline
    \end{tabular}
    }    
\end{table}

\noindent\begin{center}
		\begin{tcolorbox}[colback=black!5, colframe=black!20, width=1.0\linewidth, arc=1mm, auto outer arc, boxrule=1.5pt]                       
            {{\textbf{Answer to RQ2:} Among all three types of single software artifacts, error stack is the most effective in assisting the LLM with program repair with provided buggy method localization. Different combination of software artifacts complement each other in generating plausible patches and combining all three artifacts can achieve 56.0\% of program repair, outperforming the SOTA approaches in program repair.}}
		\end{tcolorbox}
\end{center}
\subsection{RQ3. End-to-end Performance from Code Repository to Program Repair}
\label{rq3}

Currently, most of LLM-based approaches focus on either fault localization or program repair based on the already-localized buggy methods. We investigated RQ1 and RQ2 to compare with these approaches. In contrast, our \toolName{} fully relies on LLMs for both fault localization and program repair. Furthermore, from fault localization to program repair, we not only provide the LLM with the localized buggy method signatures and lines, but also supply it with the complete method body. When the LLM is asked for the program repair task, it may reconsider and repair other parts of the code beyond the specified buggy location. Therefore, the overall end-to-end performance from the buggy code without localization to a complete program repair cannot be simply calculated by multiplying the fault localization rate by the program repair rate. Conducting a comprehensive end-to-end assessment can offer a better understanding of \toolName{}'s performance throughout the entire fault localization and program repair process.

Table~\ref{tbl:rq3-overall} shows \toolName{}'s end-to-end performance with different software artifacts. When using issue content alone, the LLM can repair 21.8\% of bugs, which is the highest among all single software artifacts. This result is also consistent with \texttt{Agentless}~\cite{xia_agentless_2024}, one of the best approaches in SWE-bench lite (with Python projects)~\cite{swe-bench-lite}, which relies solely on issue descriptions to assist LLMs in resolving GitHub issues.  

Furthermore, by combining issue content with debugging information and the error stack, the LLM can fix an additional 1.2\% and 4.0\% of  single-method bugs, respectively. When all three software artifacts are combined, the LLM can repair 28.0\% of single-method bugs and 11.2\% of non-single-method bugs. Furthermore, the combination of different artifacts achieves an end-to-end 39.7\% fix rate for single-method bugs and 17.1\% for non-single-method bugs. Figure~\ref{fig:overall_overlap} shows similar observations as other two RQs, that different combinations of software artifacts can complement each other in the end-to-end process of bug localization and program repair.
\begin{table}[!ht]
    \begin{minipage}{0.5\linewidth}
    \caption{End-to-end plausible patch in Defects4J v2.0}
    \label{tbl:rq3-overall}
    \centering
    \begin{tabular}{c|ll}
    \hline
    ~& Single Method & Non Single \\ \hline
       --- & 23/488=4.7\% & 6/346=1.7\%  \\
Issue & 103/473=21.8\% & 30/335=9.0\%  \\
Debug & 38/413=9.2\% & 3/295=1.0\%  \\
Stack & 83/488=17.0\% & 14/346=4.0\%  \\
Issue+Debug & 92/400=23.0\% & 30/283=10.6\%  \\
Issue+Stack & 123/477=25.8\% & 31/338=9.2\%  \\
Stack+Debug & 73/413=17.7\% & 13/295=4.4\%  \\
Issue+Stack+Debug & 114/407=28.0\% & 33/295=11.2\%  \\
Union & 194/489=39.7\% & 59/346=17.1\%  \\ \hline
    \end{tabular}
    \end{minipage}\hfill
    \begin{minipage}{0.4\linewidth}
        \centering
		\includegraphics[width=\linewidth,trim=270 50 10 50,clip]{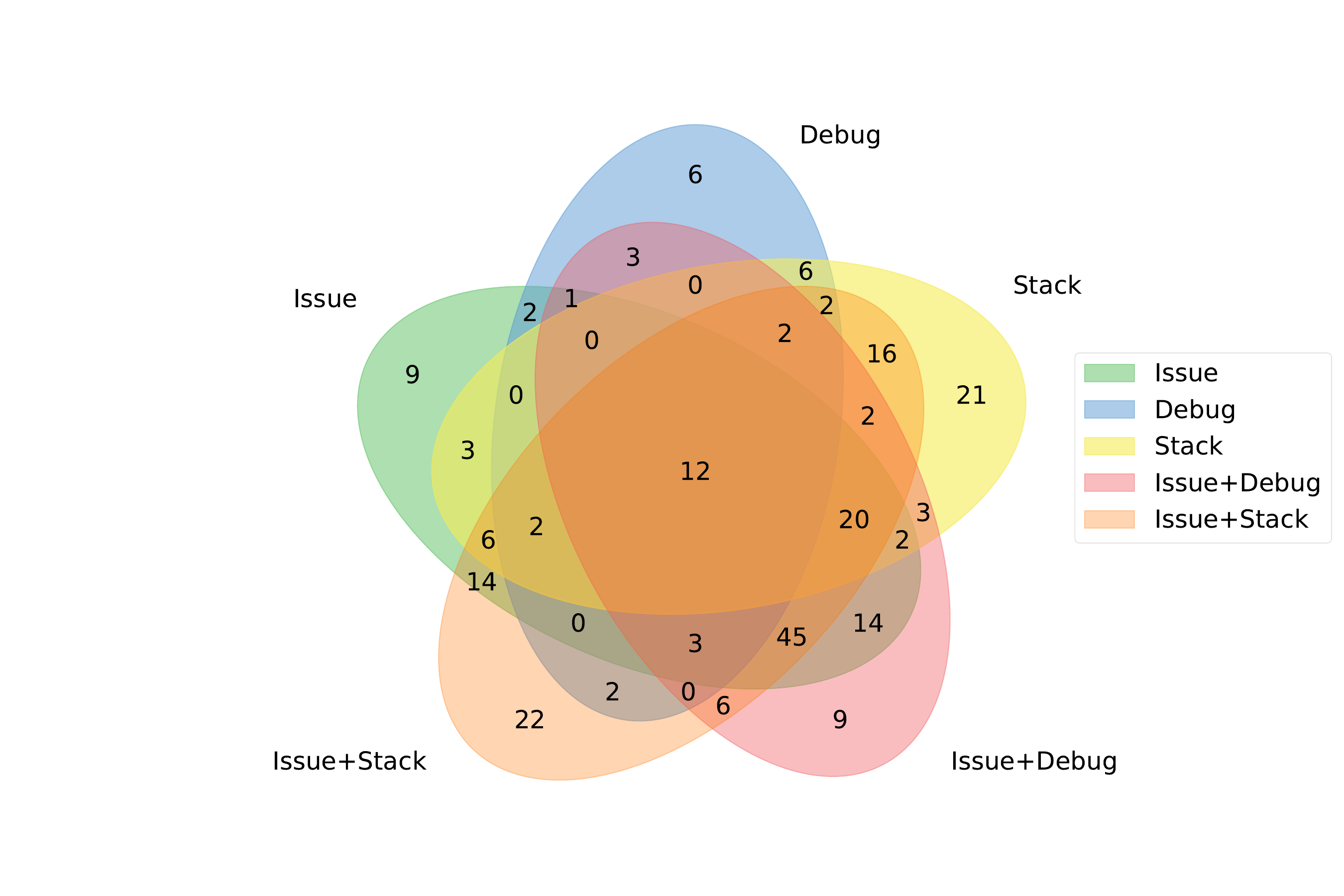}
		\captionof{figure}{Overlap of the overall performance in Defects4J v2.0}
        \label{fig:overall_overlap}
    \end{minipage}
\end{table}

\begin{table}[]
    \centering
    \caption{Time spent on each process of \toolName{}}
    \label{tbl:discussion_time}
    \begin{tabular}{cc}\hline
     Process   &  Time Spent (On Average)\\ \hline
     Extract Related Methods by \texttt{MethodRecorder} & $\approx{}$5s \\
     Localize Buggy Methods by LLM &  <1s \\
     Extract Debugging Information by \texttt{DebugRecorder} & $\approx{}$5s \\
     Localize Buggy Lines by LLM & <2s \\
     Generate Patches by LLM  & <2s\\
     Evaluate Plausible Patches by the Defects4J framework& $\approx{}$300s \\
     Total & \textbf{$\approx{}$315s} \\
     \hline
    \end{tabular}
\end{table}

We divide the time spent on each process by the number of fixed bugs to calculate the average time required to fix a bug. Table~\ref{tbl:discussion_time} shows the average time spent to fix a bug in each process of \toolName{} using our hardware device (Intel(R) Xeon(R) Platinum 8352V CPU \@ 2.10GHz, 120GB RAM). We observe that the entire localization and repair process takes approximately 15 seconds, with the evaluation of patches being the most time-consuming step, as it involves running all relevant tests. Additionally, since we select the most cost-efficient GPT-4o-mini model, the average cost to fix a bug is \textbf{\$0.057}. We calculate this by dividing the total cost of using the LLM when issue content, error stack, and debug information are provided by the number of plausible patches generated, which is 147. 


Using our \toolName{} framework for this end-to-end software maintenance activity, 253 (194 for single buggy methods and 59 for non-single buggy methods) plausible patches can be generated. To the best of our knowledge, \toolName{} outperforms all current SOTA end-to-end approaches in the Defects4J dataset (the plausible rate of \texttt{Toggle}~\cite{hossain_deep_2024} is 24.2\% for 58/240 single-hunk bugs and the plausible rate of \texttt{FixAgent} ~\cite{lee2024unified} without Web search engine is 245/835=29.3\% at a cost of \$0.364 per bug).

\noindent\begin{center}
		\begin{tcolorbox}[colback=black!5, colframe=black!20, width=1.0\linewidth, arc=1mm, auto outer arc, boxrule=1.5pt]                       
            {{\textbf{Answer to RQ3:} Among the three types of software artifacts, issue content is the most effective in assisting the LLM throughout the end-to-end process, from code repository analysis to fault localization and program repair. Different combinations of software artifacts complement each other, enhancing the overall bug localization and repair process. Moreover, our \toolName{} framework can effectively fix bugs at a low cost and within a reasonable time frame.}}
		\end{tcolorbox}
\end{center}

\section{Discussion}
\label{sec:discussion}

\subsection{Analysis of Results}

\subsubsection{A simple but efficient framework:} Unlike approaches that rely on patch skeletons~\cite{li_giantrepair_2024}, fix templates~\cite{zhang_gamma_2023}, or type checking~\cite{zhu2023tare}, \toolName{} adopts a straightforward framework that allows LLMs to handle both fault localization and program repair with the aid of two lightweight tools we implemented \texttt{MethodRecorder} and \texttt{DebugRecorder}, making it both simple and efficient. The findings of RQ1, RQ2, and RQ3 demonstrate that different software artifacts may lead to different performance in bug localization or program repair. Also, because there is no constraints such as fill-in-the-blank templates or code skeletons for generating patches~\cite{li_giantrepair_2024, zhang_gamma_2023}, \toolName{} can fix bugs that other tools cannot, especially some non-single method bugs. Our results in Table~\ref{tbl:rq3-overall} show that by combining issue content, stack error, and debug information, our approach can fix 11.2\% of non-single method bugs. More importantly, combinations of software artifacts can achieve the best performance in all experiments: bug localization, program repair, and the overall streamlined process. One explanation for the strong performance of \toolName{} is that by integrating multiple sources of information, the noise inherent in any single source can be significantly reduced~\cite{meng2024empirical}. This allows the LLM to focus on the consistent and complementary aspects of the different artifacts, enabling it to perform more effective reasoning and making better use of the available data. While it may be argued that the good results of \toolName{} are due to the use of GPT-4o models, a recent study shows that directly using the GPT-4o model does not improve the fix rate~\cite{li_giantrepair_2024}. Therefore, we argue that our proposed framework is the key to achieving the high rate in fault localization and program repair. 

\subsubsection{A strict input/output prompt design:} Table~\ref{tbl:prompt} presents the prompts used in our \toolName{} framework. In the \textsc{general task prompt}, the LLM is asked to act as a software engineer and conduct the review process, helping the LLM form a clear understanding of the overall task. The \textsc{input prompt} includes various types of software artifacts, with clear symbols denoting different hierarchy levels and structures. For example, the \textcolor{magenta}{\{related\_methods\}} in the input prompt wraps the class names in \#\#\# symbols, and the method signatures are separated by line breaks. Also, the first line of \textcolor{cyan}{\{debugging\_info\}} in the input prompt represents the currently executed method line, and the second line represents the names and values of the variables in current context (e.g., \textit{commons.lang3.math.NumberUtils:createNumber:468\{hexDigits:8\}} represents that the code is about to execute Line 468 and the value of the local variable \textit{hexDigits} is 8). These strict formatting specifications helps the LLM \leftquote{}understand\rightquote{} the structure of the various information from different software artifacts. The \textsc{expected output prompt} is very strict in \toolName{}. When localizing buggy methods, we ask the LLM to return a set of buggy method or field locations in the format \textit{path.to.ClassA::methodA}. For buggy lines, we request the LLM to return a set of buggy line locations in the format \textit{path.to.ClassA line:20}. During program repair, we employ the well-known \textit{SEARCH/REPLACE} method, which is commonly used in many state-of-the-art program repair approaches~\cite{xia_agentless_2024,ouyang2024repograph,zhang2024autocoderover,liu2024marscode}. By enforcing this strict output format, it significantly reduces the likelihood of hallucinations from the LLM. We believe that the clear and strict prompt design in our \toolName{} framework has helped the LLM achieve strong performance in fault localization and program repair.

\begin{table}[!ht]
    \centering
    \caption{A comparison from plausible patch to correct patch in Defects4J v1.2}
    \label{tbl:cmp_plausible}
    \begin{center}
    \begin{threeparttable}
    \begin{tabular}{c|cccc}\hline
         Project & \toolName{} & GaintRepair~\cite{li_giantrepair_2024} & Tare~\cite{zhu2023tare} & TBar~\cite{template:liu_tbar_2019}  \\ \hline
         Chart &    7/12 &7/10 &11/14&7/10  \\
         Closure &  10/16 &16/33&12/23&6/10  \\
         Lang &     8/16 &12/19&12/19&4/11 \\
         Math &     28/48 &22/40&18/34& 12/26 \\
         Time  &    0/2 &1/3  &2/3  & 1/2 \\
         Mockito &  6/11 &6/6  &2/2  & 1/2 \\
         Total   &  59/105 &64/111&57/95&31/61  \\ \hline
         P(\%)   &  56.2\% &57.7\%&60.0\%&50.8\%  \\ \hline
    \end{tabular}
    \begin{tablenotes}
    \item[*] X/Y denotes X correct patches and Y plausible patches.
    \end{tablenotes}
    \end{threeparttable}
    \end{center}
\end{table}

\subsubsection{Plausible patch to correct patch:} Most state-of-the-art (SOTA) program repair approaches use plausible patches that pass all unit tests as an important evaluation metric, since generating plausible patches within limited time and resources is crucial for practical applications. To facilitate a better comparison, we also used plausible patches in RQ2 (Which software artifacts can better assist LLMs in generating valid patches when provided with buggy methods?) and RQ3 (What is the overall performance of \toolName{} in bug localization and program repair?).  Some may argue that a high number of plausible patches does not necessarily translate to a high number of semantically correct patches. To address this, we manually inspected the plausible patches, using the approach described in Section~\ref{patch_validation}. Table~\ref{tbl:cmp_plausible} compares the end-to-end repair results on Defects4J v1.2, as all the baselines consistently used this benchmark. Our \toolName{} framework uses the LLM to localize buggy methods and lines, whereas all three approaches — \texttt{GiantRepair}~\cite{li_giantrepair_2024}, \texttt{Tare}~\cite{zhu2023tare}, and \texttt{TBar}~\cite{template:liu_tbar_2019} — employ the spectrum-based algorithm Ochiai~\cite{ochiai}, implemented by GZoltar~\cite{gzoltar}, to localize buggy methods. As shown in Table~\ref{tbl:cmp_plausible}, the ratio of correct to plausible patches (56.2\%) produced by our \toolName{} is lower than that from Tare (60.0\%) and from \texttt{GiantRepair} (57.7\%). This difference is understandable, as \texttt{GiantRepair} enforces an AST patch skeleton and \texttt{Tare} uses a typing-checking mechanism. Notably, 59 out of 105 plausible patches generated by \toolName{} are evaluated as correct patches, which is higher than the  57 in \texttt{Tare} and 31 in \texttt{TBar}. This finding demonstrates \toolName{}'s ability to generate candidate patches that not only pass all unit tests but are also semantically correct.

\subsection{\color{black}{Comparison with Baselines of Different Models or on Different Languages}}

\textcolor{black}{\subsubsection{Comparison with Baselines of Different Models} We selected state-of-the-art LLM-based fault localization and program repair frameworks on the \dforj{} dataset as our baselines. One may argue that some baselines do not use a GPT-4o model, and that the superior performance of our \toolName{} framework is due to the use of a more powerful model. However, as shown in Table~\ref{tbl:compare_FL_sota}, even using the same GPT-3.5 model as the baselines, our framework outperforms existing state-of-the-art fault localization approaches. For program repair, the models used in these baselines vary significantly. For example, \texttt{GiantRepair} and \texttt{Repilot} combine the outputs of multiple LLMs, while \texttt{FitRepair} fine-tunes CodeT5 using 50\% of the \dforj{} dataset. As a result, it is challenging to align all baselines under a single model. According to a recent study~\cite{campos2025empirical}, GPT-4o achieves only 16.4\% and 27.8\% accuracy at pass@1 and pass@5, respectively, in generating correct patches given a buggy function on the \dforj{} dataset. Additionally, as noted in the discussion section of the GiantRepair paper~\cite{li_giantrepair_2024}, given a buggy function, GPT-4o-mini achieves 21.7\% accuracy when generating 200 patches per bug on the \dforj{} dataset (pass@200).  In contrast, \toolName{}, which leverages issue content, error stack and debug information, achieves 56.0\% accuracy in program repair using GPT-4o-mini (pass@30). We believe that this performance gap is due to our framework’s ability to effectively leverage various software artifacts.}  

\textcolor{black}{\subsubsection{Comparison with Baselines on Different Programming Languages} Our framework was largely inspired by \texttt{Agentless}. Initially, we attempted to apply \texttt{Agentless} to the \dforj{} dataset, but the resolution rate was below 10\%. This led us to consider incorporating debugging information and re-implementing \texttt{Agentless} for evaluation on the \dforj{} dataset. During this process, we encountered a major issue: \texttt{Agentless} uses only the skeleton of the source files during the fault localization step, without incorporating the detailed source code. To address this limitation and incorporate both debugging information and actual code lines, we extended the original \texttt{Agentless} framework to include the full source code as input to the LLM. However, this introduced a new challenge: the input frequently exceeded the LLM’s token limit. To overcome this, we introduced \texttt{MethodRecorder}, which identifies and retains only the methods relevant to the failing test cases, thereby reducing the number of code elements passed to the LLM. We also pruned the debug information by recording only variables that had changed and cropping long arrays and lists to further reduce the token size. Without such adjustments or modifications to \texttt{Agentless}—even when providing all the artifacts listed in this paper—its end-to-end performance remained below 15\%. This demonstrates that directly applying a leading program repair framework designed for Python code to Java code does not yield competitive results.}

\begin{table}[]
    \centering
    \footnotesize
    \caption{\textcolor{black}{Resolution rates of frameworks using a GPT-4o model on the SWE-bench Lite Leaderboards}}
    \label{tbl:swe}
    {\color{black}\begin{tabular}{cccc}
    \toprule
    \textbf{Tool} & \textbf{LLM} & \textbf{Resolved/Selected} & \textbf{\%Resolved} \\ \midrule
    \toolName{}  & GPT-4o-mini& 46/207 & 22.2\% \\ \hline
    SWE-agent~\cite{swe-agent}    & GPT-4o& 55/300 & 18.3\% \\
    AppMap Navie~\cite{appmap} & GPT-4o& 65/300 & 21.7\% \\
    Aider~\cite{aider}        & GPT-4o+Claude-3 & 79/300 & 26.3\% \\
    Agentless~\cite{xia_agentless_2024}    & GPT-4o& 82/300 & 27.3\% \\
    MarsCode~\cite{liu2024marscode} & Multiple LLMs(includes GPT-4o) & 118/300 & 39.3\% \\
    \bottomrule
    \end{tabular}}
\end{table}

\begin{figure}[!tbp]
    \centering
    \includegraphics[width=0.5\textwidth,trim=200 20 0 20,clip]{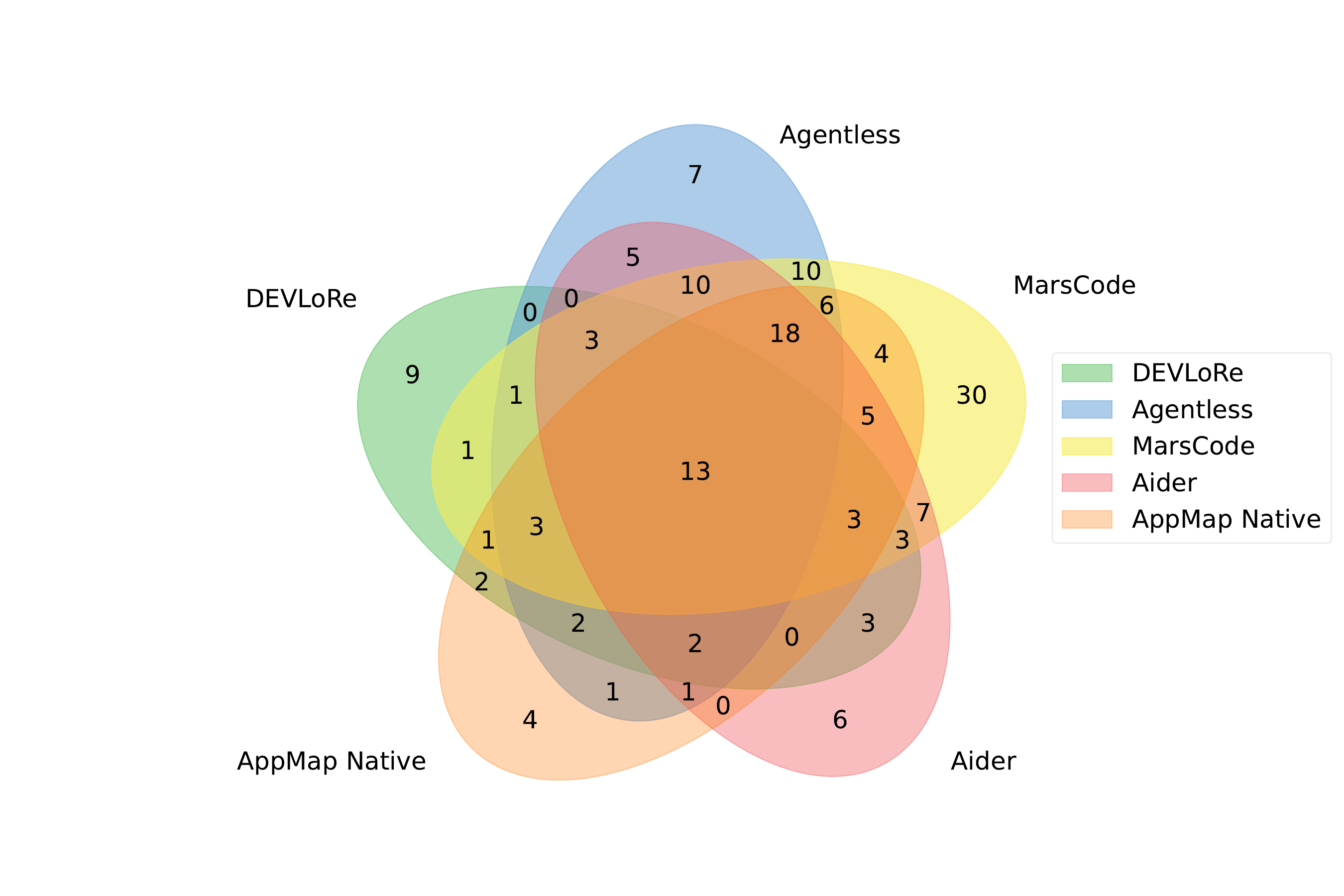}
    \caption{\textcolor{black}{A comparison between \toolName{} with baselines on SWE-bench Lite}}
    \label{fig:overlap_swe}
\end{figure}

\textcolor{black}{To compare with leading frameworks and evaluate \toolName{}’s capability on Python code, we re-implemented our framework and tested it on SWE-bench Lite. The first challenge we encountered was that, even after filtering with \texttt{MethodRecorder}, the length of the related methods for one-third of the bugs in SWE-bench Lite still exceeded the LLM’s token limit. As a result, we were only able to provide related methods to the LLM for 207 out of 300 bugs in SWE-bench Lite. Additionally, we observed that Python’s debugging information is significantly more verbose than that of Java. For some bugs, the required patches involve modifying top-level constructs such as imports or global variables, for which debugging information cannot be extracted by executing the test cases. As shown in Table~\ref{tbl:swe}, among the 207 bugs with related methods information, \toolName{} successfully fixed 46 (46/207=22.2\%). We also provide a comparison with the SWE-bench Lite Leaderboards using the GPT-4o model. The results in Table~\ref{tbl:swe} shows that \toolName{} did not outperform \texttt{Agentless} or \texttt{SWE-agent} or \texttt{MarsCode} in the total number of fixed Python bugs. We suspect this is due to fundamental differences in coding practices between Java and Python, which result in related methods or debugging information being unavailable for a portion of the issues. Moreover, we compared the resolved issue IDs of \toolName{} with those of baselines. The overlap of resolved issues among the baselines and \toolName{} in Figure~\ref{fig:overlap_swe} shows that \toolName{} was able to resolve 9 unique issues, despite resolving only 46 issues in total. Manual inspection of these 9 issues revealed that all of them involved the use of debugging information. This result suggests that the debugging information automatically extracted in our framework—among these baselines, only \texttt{AppMap Naive} utilizes runtime data when available in an interactive way—shows promise in addressing a subset of issues. Moreover, this experiment highlights that directly applying a leading framework from SWE-bench to \dforj{}, or vice versa, may not yield state-of-the-art results. Generalizing the framework across different programming languages will be part of our future work.}

\subsection{Threats to Validity}
The first threat to validity is the accuracy of the two tools we implemented: \texttt{MethodRecorder} and \texttt{DebugRecorder}. Both tools rely on mature Java agent technology~\cite{java8instrumentation}. We randomly selected several projects and manually verified the outputs of both tools. The manual inspection showed that the outputs were accurate. However, we did not inspect all projects, which may pose a threat to the construct validity.

Another potential threat is that the GPT-4o-mini model used in this work may have been trained on open-source projects from GitHub, which could overlap with the \dforj{} dataset, leading to possible data leakage. To mitigate this, we also randomly selected 100 bugs from another dataset, GrowingBugs~\cite{jiang2022bugbuilder}, and found the plausible fixing rate to be 39\%, which may help alleviate this concern. Additionally, the debugging information requires dynamic analysis, which is unlikely to have been used during the model’s training. Also as found in \cite{li_giantrepair_2024,campos2025empirical}, directly using the GPT-4o model can not improve the fix rate. 

\textcolor{black}{The final threat to external validity is that our systematic experiments used only one model and one approach, although using GPT-3.5 in \toolName{} also outperform SOTA fault localization approaches. But the findings may not generalize to other models or approaches. We plan to explore this in our future work. Also, as discussed in Section 6.2, generalizing our framework to other programming languages, such as Python, does not currently achieve state-of-the-art performance. We plan to further improve and evaluate \toolName{} on additional datasets across multiple programming languages in our future work.}
\section{Related Work}
\label{sec:related}

\subsection{Large Language Models for Fault Localization}
Recently, there has been significant interest in using LLMs for fault localization. \texttt{Toggle} incorporated additional contextual information, such as the buggy line number or code review comments, and greatly enhances the accuracy of predicting both the starting and ending buggy tokens~\cite{hossain_deep_2024}. \texttt{AGENTFL} employs a multi-agent system based on ChatGPT and frames the fault localization task as a three-step process: comprehension, navigation, and confirmation. In each step, \texttt{AGENTFL} deploys specialized agents, each with unique expertise, and uses different tools to address specific tasks~\cite{soapfl}. \texttt{CrashTracker} conducts static analysis to map each crash to the corresponding exception instance and identify potential buggy candidates. It then utilizes LLMs to enhance the explainability of the localization results~\cite{yan_better_2024}. Jiang et al. assessed the performance of recent commercial closed-source general-purpose LLMs, such as ChatGPT 3.5, ERNIE Bot 3.5, and IFlytek Spark 2.0, on line-level fault localization with the provided buggy method~\cite{jiang_evaluating_2024}. \texttt{LLMAO} fine-tunes LLMs with 350M, 6B, and 16B parameters on small, curated corpora like Defects4J, improving Top-1 fault localization by 2.3\%-54.4\% and Top-5 results by 14.4\%-35.6\%, compared to the state-of-the-art machine learning fault localization~\cite{yang_large_2024}. \texttt{AutoFL} prompts an LLM to use function calls for navigating a repository, enabling effective fault localization in large codebases while overcoming the LLM context length limit. It also generates an explanation of the bug and suggests a fault location~\cite{autofl}. \texttt{LLM4FL} combines traditional spectrum-based fault localization with prompt chaining to divide large coverage data into manageable groups. By employing multiple LLM agents, it navigates the codebase more effectively to localize faults~\cite{rafi_enhancing_2024}. Like \texttt{LLM4FL} and \texttt{AutoFL} which combines transitional fault localization tools, \toolName{} incorporates a static code analysis tool, \texttt{MethodRecorder}, to identify relevant buggy methods in a lightweight manner by invoking failing tests. However, unlike the above approaches, we provide the LLM with a combination of software artifacts, including issue content, error stack traces, and debugging information, which are commonly used by human developers for fault localization. This enables \toolName{} to outperform state-of-the-art fault localization methods with greater efficiency and lower cost.

\subsection{Large Language Models for Program Repair}
Recent studies have extensively explored the use of LLMs for program repair. \texttt{ChatRepair} initially provides the LLM with relevant test failure information and then learns from both the failures and successes of previous patching attempts for the same bug, enhancing its ability for more effective APR~\cite{chatrepair}. \texttt{GAMMA} converts various fix templates into mask patterns and leverages a pre-trained language model to predict the correct code for the masked portions, treating APR as a fill-in-the-blank task~\cite{zhang_gamma_2023}. \texttt{Repilot} generates a candidate patch by combining LLM suggestions with a Completion Engine, removing infeasible tokens and filling in gaps proactively~\cite{repilot}. \texttt{FitRepair} integrates the plastic surgery hypothesis into LLM-based APR, combining the direct use of LLMs with two domain-specific fine-tuning strategies and one prompting strategy to enhance its repair capabilities. It can directly generate the correct code in context, effectively ``filling in the blanks'' of missing code lines or hunks~\cite{fitrepair}. \texttt{Tare} incorporated type checking into neural program repair model and can successfully repair 62 and 32 bugs from Defects4J v1.2 and Defects4J v2.0~\cite{zhu2023tare}. \texttt{GiantRepair} creates patch skeletons from LLM-generated patches to narrow the patch space, then generates context-aware, high-quality patches by instantiating these skeletons for specific programs~\cite{li_giantrepair_2024}. \texttt{FixAgent} unifies debugging through multi-agent collaboration and achieves strong performance with a three-layer hierarchical structure, where the final layer involves the use of a Web search engine~\cite{lee2024unified}. \texttt{MORepair} fine-tunes LLMs for program repair by adapting both to the syntactic nuances of code transformation and the underlying logic of code changes, enabling the generation of high-quality patches~\cite{yang2024multi}. To leverage LLMs' capabilities and augmented information, \texttt{CREF} is a semi-automatic repair framework for programming tutors, highlighting the potential for enhancing LLMs’ repair capabilities through tutor interactions and historical conversations~\cite{yang2024cref}. \texttt{RepairLLaMA} is an innovative program repair method that finds optimal code representations for APR using fine-tuned models, and introduces a state-of-the-art parameter-efficient fine-tuning technique (PEFT) for program repair~\cite{silva2023repairllama}. Our approach differs from the aforementioned approaches in two key ways. First, we do not rely on fill-in-the-blank templates or skeletons for generating patches, which helps avoid patch overfitting problems in program repair~\cite{smith2015cure,le2018overfitting,feipatch}. Second, our fault localization step allows for the identification of multiple buggy methods, which are then fed into the LLM's repair process along with various software artifacts. This provides flexibility to address not only single-method bugs but also bugs spanning across different methods. As seen in Table~\ref{tbl:rq3-overall}, the end-to-end process from localization to repair can generate plausible patches for 17.1\% of non-single-method bugs, while most existing program repair approaches primarily target single-method bugs. \textcolor{black}{A recent study~\cite{parasaram2024fact} investigated how to identify an optimal set of facts to maximize the performance of LLM-based program repair. Their framework adopts chain-of-thought prompts, uses artifacts from both the fixed (correct) and buggy programs, and is evaluated on the BugsInPy dataset. Complementary to their work, we show that extracting all artifacts solely from the buggy version—which can be found in most development scenarios—and using direct prompts can still yield effective results on both the Defects4J dataset and SWE-bench Lite. Moreover, our paper demonstrates that different combinations of artifacts can enhance one another, helping LLMs on both fault localization and program repair.} 

\section{Conclusions and Future Work}
\label{sec:conclusion}

This paper presents an LLM-based framework, \toolName{}, for streamlining fault localization and program repair. By mimicking human developers in addressing bug problems and integrating three different software artifacts, \toolName{} demonstrates strong performance in both fault localization and program repair, outperforming current state-of-the-art approaches in terms of bug fixing rate, time, and cost on the Defects4J dataset. In addition, unlike more rigid approaches that ask LLMs to fill in the blank within a single buggy method or use the fix template, \toolName{} feed the LLM with only different software artifacts and there are no constraints on the \toolName{} framework regarding how it repairs bugs, \toolName{} has shown significant potential in handling bugs that span across multiple methods.

Our future work will focus on the following directions: First, expanding \toolName{} to support additional programming languages such as Python and C/C++, and evaluating and improving its effectiveness on projects written in these languages. Second, testing the \toolName{} framework with a broader range of software artifacts—such as commit history, code review comments, and user documentation—to assess how these additional data sources can further enhance the bug localization and program repair process. \textcolor{black}{Third, conducting a systematic evaluation involving multiple approaches and LLMs to better assess \toolName{}'s generalization capabilities.}

\section*{Data Availability}
The source code and experimental results of this work for replication are available at \url{https://github.com/XYZboom/DEVLoRe}.

\begin{acks}
This work has been partially supported by the National Natural Science Foundation of China (NSFC) with Grant No. 62172311 and the Major Science and Technology Project of Hubei Province under Grant No. 2024BAA008. The numerical calculations in this paper have been done on the supercomputing system in the Supercomputing Center of Wuhan University.
\end{acks}

\bibliographystyle{ACM-Reference-Format}
\bibliography{ref}

\end{document}